\documentclass[a4paper,12pt,onecolumn,draftclsnofoot]{IEEEtran}
\usepackage{amssymb}
\usepackage{tabularx}
\usepackage{graphicx}
\usepackage{caption}
\usepackage{subcaption}
\usepackage[tbtags]{amsmath}
\usepackage{amsfonts}
\usepackage{mathrsfs}
\usepackage{multirow}
\usepackage[english]{babel}
\usepackage{cite}
\usepackage{url}
\usepackage{hyperref}
\hypersetup{
  colorlinks=true,
  linkcolor=red,
  linktoc=all,
  citecolor=blue,
  bookmarksnumbered=true
}
\usepackage{cleveref}
\usepackage{units}
\usepackage[section]{placeins}

\usepackage[colorinlistoftodos]{todonotes}

\IEEEoverridecommandlockouts

\begin{document}
\title{COVID-19 Epidemic in Mumbai: \\
 Projections,
full economic opening,  \\
and containment zones versus contact tracing and testing: An Update }
\author{
\IEEEauthorblockN{
TIFR Covid-19 City-Scale Simulation Team \\
Prahladh Harsha,
Sandeep Juneja,
Daksh Mittal,
Ramprasad Saptharishi
\\}
{October 29, 2020} \\
}
\maketitle

{\let\thefootnote\relax
  \footnotetext{
    Source code available at \url{https://github.com/dasarpmar/epidemics-simulator-mumbai/releases/tag/v4.0}
  }
}
\vspace*{-2.5cm}
\section{Summary}

Mumbai, amongst the most crowded cities in the world, has witnessed the fourth
largest number of cases and the largest number of deaths among all the cities in
India.  The first case in Mumbai was detected on 11 March 2020, and the first
fatality was recorded on 17 March 2020. Currently, as of 26 October, 2020, Mumbai
has reported 252,087 cases and 10,099 fatalities, thus contributing
disproportionate share to India's tally of 7.94 million reported cases and
119,014 deaths. Mumbai, along with the rest of India has been in a lockdown since
March 25, 2020. Initially imposed for three weeks, this lockdown was extended in
Mumbai and other parts of India till May 17, 2020. Thereafter, Mumbai has seen
gradual relaxations in population movement.  In particular, in the 
``Mission Begin Again'' order dated 31st August
2020~\cite{maharashtra20:_easin}, the Government of Maharashtra has allowed 20\%
attendance in workplaces.  

As per the release by the government, the Indian economy contracted by
23.9\% in the first quarter of the fiscal year 2020-21.  Given the
large economic toll on the country from the lockdown and the related
restrictions on mobility of people and goods, swift opening of the
economy especially in a financial hub such as Mumbai becomes
critical. However, opening up of Mumbai is crucially linked to opening
its crowded public transit systems, especially the crowded suburban
trains. Too swift an opening may lead to a sudden increase in the
spread of the epidemic leading to a `difficult to manage' second wave
of hospitalisations.
 
Fortunately, the curve of medical indicators for Mumbai such as
hospitalisations, critical patients, reported cases and fatalities at
any time, had begun to stabilize or `flatten' over the months of June
and July and further reduced in August. 
 There was some increase in medical indicators  from late 
August onwards which appears again to be stabilizing by mid-October.
This was very likely due to the increased intermingling due to the Ganpati festival combined with the opening up
of the economy.
 Mumbai Sero-Survey\cite{Mumbai_sero2020} indicated high prevalence in the city
 and particularly in the slums in early July. Given the increase in reported cases from the city
 especially from the non-slum areas since then, the overall prevalence in Mumbai is likely
 to be quite high, thus  allowing the city  room to further open up. In this
 report, we use our IISc-TIFR agent based simulator described in
detail in \cite{City_Simulator_IISc_TIFR_2020} to develop long term
projections for Mumbai under realistic scenarios related to Mumbai's
opening of the workplaces, or equivalently, the economy, and the
associated public transportation through local trains and buses.
 
These projections were developed taking into account a possible second
wave if the economy and the local trains are fully opened either on November 1, 2020 or on January 1, 2021. 
 The impact on infection spread in
Mumbai if the schools and colleges open on January first week 2021 is
also considered.  We also capture the increased intermingling amongst the population during the Ganpati festival
as well as around the Navratri/Dussehra and Diwali festival.
Our conclusion, based on our simulations, is that the
impact of fully opening up the economy on November 1 is
manageable provided reasonable  medical infrastructure is in place. Further, schools and
colleges opening in January do not lead to excessive increase in
infections. While Ganpati festival had a substantial impact on medical indicators, if the intermingling level
during Navratri/Dussehra and Diwali festival is similar, then, since these festivals occur later when a larger fraction of the population has already been infected,  the resulting infections are likely to be less,
and thus the overall impact on city's medical infrastructure also relatively less. Though not explicitly modelled, we expect similar conclusions
to hold for the Chat festival and Christmas later in the year.
Of course, if there is a substantial increase in interaction amongst the population during the upcoming festivals
and the social distancing/masks related precautions are relatively weakened, then one may again
see a significant rise in infections. 

\subsubsection*{Changes from the earlier report \cite{harsha2020covidmumbai}.}
This report is an update of an earlier one we released in early September\cite{harsha2020covidmumbai}. The key changes, and the explanation for the modifications,  are given below:
\begin{itemize}
\item To improve the modelling of the infection in high-density areas (slums) to better match the  observed prevalence data \cite{Mumbai_sero2020}, we increase the high-density transmission rate (\texttt{HD-FACTOR} set to $3$, instead of $2$).
\item We use the data from \cite{malani2020serosurvey} for age distribution in high-density areas and low-density areas  to capture the fact that the high-density areas have a younger population (see Figure~\ref{Age-Distribution}) . This results in a lower fatality rate per infection within the high-density areas, and also an increase in spread within the high-density areas since there are more individuals in the working age-group. 
\item The population of Mumbai is taken to be 12.8 million as opposed to 12.4 million.
  This in consonance with mid-year estimated population 2019 data with MCGM that updates the 2011 census data \cite{mumbai_census2011_A,suburban_mumbai_census2011_A}.
\item The intervention modelling includes lower compliance and higher community level interactions during the festive seasons. The details of this are explained in \autoref{sec:scenarios}. 
 \end{itemize}

  The net result of these changes is that in our model slums see high infections early on
largely during April to June. Infections thereafter are primarily from non-slums. This matches 
the Mumbai scenario much better.
 Later in this report we also conduct some rudimentary experiments to ball park gauge the benefits and the costs of
  introducing {\em perfect} vaccines amongst the older population of Mumbai.

\bigskip

Our simulations further suggest that by mid-January  2021, 
the prevalence (fraction of the population infected) can
be seen to be stabilising close to 80\% in slums and 55\% in
non-slums. This stabilisation and  high prevalence indicates that Mumbai city may have
more or less reached ``herd immunity'' by then. By this we mean that the new
infections and related medical indicators 
in the city will be substantially reduced compared to their peak values
in mid-May and June 2020.

In these simulations we also conduct counterfactual experiments where
the containment efforts as well as the contact tracing and testing
efforts are varied and their impact on the health indicators is
measured. Our simulations suggest that containment efforts do a better
job of slowing infection spread compared to increasing the contact
tracing and testing efforts. Increase in the latter leads to only
marginal improvement in slowing the infection.

 Below, we list some policy recommendations for opening up the workplaces and schools and colleges  in Mumbai that incorporates above considerations.

 \subsection{Policy Recommendations}
 
 \begin{itemize}
 \item We had in the earlier version of this report \cite{harsha2020covidmumbai}, recommended gradual opening of the workplaces so that the
   increase in infections that may result from increased occupancy in
   local trains and other public transport is manageable. 
   We continue to recommend similar gradual opening up of the city
   and that the economy may be fully opened up
   by November 1 or soon thereafter, again  with the opening up process carefully 
   based on observed infections. 
   \item
    Schools and colleges
   may be opened by first week of January 2021.  As mentioned earlier,
   our simulations suggest that the resulting second wave from this
   opening up is minimal.
\item 
 Social distancing in public transport, staggering of
  office times, use of shifts to the extent feasible is recommended.  
\item
  Prevailing hygiene measures such as mandatory use of masks/face-covers,
  encouragement of regular hand-hygiene, regular disinfection of ``high-touch
  surfaces'' in trains and workplaces~\cite{who20:_q} etc.~should continue as
  before.
\item Our analysis of containment zones vis-a-vis contact tracing and testing
  suggests that wherever feasible and when the economic costs are not
  prohibitive, containment in regions where infection is seen to be present is a
  desirable option to slow the infection spread.
\end{itemize}

In addition to the various modeling assumptions listed in our previous
reports~\cite{City_Simulator_IISc_TIFR_2020,IISc_TIFR_2020_Mumbai_Report2}, our
recommendations rely on two important assumptions.
\begin{enumerate}
\item Our first assumption is that the population, by and large, will continue
  to observe social distancing precautions including wearing of masks. This
  could change as public perception of risk changes over time. This may lead to
  increase in infections not accounted for in our projections.
\item Our second assumption is that the reinfection probability is sufficiently
  small for that population of Mumbai that it can be ignored in opening up of
  the city. We note that cases of reinfection have recently been
  reported. However, the number of such reports continues to be very few. If
  this changes and reinfection happens to a non-negligible proportion of the
  population, then our projections become less valid.  
\end{enumerate}

\bigskip

{\noindent \bf Vaccines:} It is believed that a COVID-19 vaccine will become available sometime in 2021. 
Important decisions related to prioritising
people to vaccinate and to manage the storage and distribution of vaccines
would require careful analysis. 
This analysis would also include the cost and time taken in vaccine production, 
the effectiveness of the vaccines,
the frequency of administering them to each person,
the time that a vaccine takes to provide immunity, and the duration
for which the immunity is provided. These are complex issues
that require detailed analysis. Later in this report we conduct some very rudimentary  simulation experiments related to administering vaccines
under simplified assumptions to get a ball park idea of the potential costs and benefits,
in terms of number of fatalities and the load on medical facilities, 
of administering vaccines to relatively older population in Mumbai.

Our broad conclusions are that 
Greater Mumbai has an estimated 13.1 lakh people aged 60 and older.
If these are all vaccinated by February 1 with a vaccine that provides instantaneous and perfect immunity,
then the number of fatalities post February 1 will reduce by estimated 53\% from around 950 to 450 in the next six months.
The hospitalisations (including critical cases) will reduce by estimated 40\% from around 8840 to 5340
in the next six months.
Similarly, if estimated 29.3 lakh  50 years and older Greater Mumbai residents
are vaccinated on February 1, then 
the number of fatalities post February 1 will reduce by estimated 64\% from around 950 to 340  in the next six months.
The hospitalisations (including critical cases) will reduce by estimated 67\% from around 8840 to 2910
in the next six months.

\bigskip

There have been varying news-reports on the differences among the
population in how the disease spreads with respect to age, especially
among the younger population. In particular, the role of the younger
population in the transmission of the disease is not
well-understood. In our earlier report~\cite{harsha2020covidmumbai},
we took a  cautious view and recommended opening of schools
and colleges only as late as January 1, 2021, several months after the
opening of the economy. Recent studies in the Indian states of Andhra
Pradesh and Tamil Nadu~\cite{Laxminarayan2020} suggest that children in
the age group of 0--14 pose a significant transmission risk,
corroborating our recommendation.

\bigskip

 {\bf Caveats:}
We emphasize that our simulator is intended primarily as a tool for comparing the effectiveness of different non-medical interventions to assist decision making. In particular, the simulator, due to the inherent model uncertainty, is not intended as a tool for predicting absolute numerical values of COVID-19 cases.  In our informal view (which is difficult to validate scientifically),  a confidence interval of $\pm$  20\% to  30\%  around the projected numbers may be  reasonable to capture the model uncertainty. This number may be larger when estimates with small values are considered.
On the other hand, the statistical error due to the random noise in the  simulations is much smaller and is easily controlled.

We also recognize that many of the non-pharmaceutical interventions considered in our study, especially when they remain implemented over a long duration, may lead to important social and economic concerns and consequences, beyond their effect on the evolution of the epidemic. Prolonged restrictions, besides economic concerns, may also lead to disruption of many essential supply chains and medical services. The World Health Organisation Pulse report \cite{WHO-pulse}, based on responses from over 100 countries, show significant reductions in routine vaccinations, diagnosis and treatment of noncommunicable diseases, antenatal care, cancer diagnosis and treatment, and many others. These are important factors that must be taken into consideration when implementing any prolonged restrictions. However, we do not know how to quantitively project many of the socio-economic effects of a non-pharmaceutical intervention. The scope of our simulator is to project purely COVID-19 related stresses on the medical infrastructure. The modelling of such socio-economic effects, though important,  remains beyond the scope of our simulator.

Similar to our previous report, we emphasize that this report has been prepared to help researchers and public health officials understand the effectiveness of social distancing interventions related to COVID-19 in terms of the stresses on the medical infrastructure. The report should not be used for medical diagnostic, prognostic or treatment purposes or for guidance on personal travel plans.

\pagebreak[2]

\section{Towards fully opening  Mumbai}

Greater Mumbai (consisting of Mumbai and Suburban Mumbai) has a population of about 1.28 crores (12.8 million)  and a population density of roughly 21,000 per km$^2$  making it one of the densest cities in the world\footnote{Some of the discussion in the Introduction first appeared in
\cite{IISc_TIFR_2020_Mumbai_Report2}}.  Further, about 53\%~\cite{MCGM_census2011_report}   of Mumbai lives in cramped dwellings with shared sanitation facilities where the population density may be 5 to 10 times larger than other parts of the city.  In addition, crowded suburban trains are the lifeline of the city where the suburban railway system serves more than 80 lakh (8 million) passengers on a weekday, in normal times \cite{railway-survey}. It is generally believed that the infection spreads faster in denser areas, due to increased contacts in these areas. Given these factors, the public health threat in Mumbai is particularly 
acute. The importance of modelling the effect of infection spread arising from the gradual opening and relaxation of lockdown
 measures, for a 
city like Mumbai, cannot then be over-emphasized. We model the spread of infection in the city using our IISc-TIFR agent-based city simulator \cite{City_Simulator_IISc_TIFR_2020}. For completeness, we briefly review it below.  

\vspace*{.1in}

{\noindent \bf Agent-based city simulator (ABCS):} As described in detail in \cite{City_Simulator_IISc_TIFR_2020}, our agent-based simulator creates a synthetic model of about  1.28 crore (12.8 million) residents of Mumbai that matches the city population ward-wise, and matches the numbers employed, numbers in schools, commute distances, etc. This is done by suitably populating households, schools, and workplaces with people. Several interaction spaces including households, local communities, schools, workplaces, trains, etc. are then modelled to realistically capture the spread of infection. The synthetic city is then seeded with infections to match the observed fatalities till April 10. The infective individuals expose the susceptible individuals to the disease through their interactions in the various interaction spaces. The disease then incrementally evolves in time. The tool helps keep track of the number infected in the city as well as the disease progression within an infected individual. A person infected by the disease may remain asymptomatic and recover, or may develop symptoms. A symptomatic person may recover or may develop severe symptoms and be hospitalised. A patient hospitalised may recover or may become critical. A critical patient may recover or may become deceased. The disease progression parameters are based on \cite{verity2020estimates}.

\subsection{Scenarios considered}\label{sec:scenarios}

In this work we report the following scenarios:
\begin{itemize}
\item
{\bf Long term forecasts:}  We  develop long term forecasts till March 15, 2021 under the following six scenarios:
\begin{itemize}
\item
Containment effort set at 75\% and at 60\%. Exact modelling of containment effort relies on the modelling feature `neighbourhood containment zones'  introduced in \cite{City_Simulator_IISc_TIFR_2020} and is discussed later
in Section~III. 
\item
Train infection levels are set at normal $\beta_T=0.19 \times \beta_H$  (see
\cite{IISc_TIFR_2020_Mumbai_Report2}  for detailed calculations to arrive at this number; that report also discusses the household transmission parameter
$\beta_H$ and the rationale that relates it to $\beta_T$. The value of $\beta_H$ used is calibrated primarily to fatality data, and is given in
Figure~\ref{fig:beta-values}.) as well as to the larger value of
 $\beta_T =0.30 \times \beta_H$ to account for the additional infections that may occur in trains
 through passengers coming to Greater Mumbai 
 through the neighbouring areas in the Mumbai Metropolitan Region (MMR).
 We further consider a more pessimistic higher value of
  $\beta_T =0.40 \times \beta_H$ to account for the inherent uncertainty 
  in estimating $\beta_T$ without availability of 
  relevant data on infection spread through trains.
\end{itemize}

 The workplace attendance is a good measure of economic activity. It is set in our model as follows:
Lockdown till May 17. Mobility to workplaces set at 5\% from May 18 to May 31. In June this 
is set at 15\% and it increases 
to 25\% in July. It is set at level 33\% in August. For September and October 
it is set at 50\%. And thereafter it fully opens (100\% attendance) from November onwards.

Festivals in Mumbai are a time for increased intermingling amongst the population, and during this time
compliance on wearing masks, maintaining social distance, etc. is likely to be less. We account for this in our model 
as follows:
\begin{itemize}
\item
To account for increased intermingling due to the Ganpati festival, from August 20 to September 1,
we increase $\beta$ for community by 2/3, and we reduce compliance from 60\% in non-slums and 40\% in slums to
40\% in non-slums and 20\% in slums. With this adjustment we observe that the fatality data from the model
is fairly close to the actual observed fatalities in months of September and October.  
\item
We do a similar adjustments for Navaratri and Dussehra for the period October 19 to October 25. For Diwali
we conduct similar adjustments for the period November 8 to November 14. 
\end{itemize}

The developed model is validated by comparing the model projections with the observed health data, that is,  observed number of fatalities, hospitalisations and critical cases. 

As in \cite{IISc_TIFR_2020_Mumbai_Report2}, in all our simulations we continue to assume that outside the festival times,
60\% of households are compliant in residential, relatively low density areas (non-slums), while 40\% 
of households are compliant in slums or high density areas. 

\item
{\bf Fully operational economy:} We consider the following three scenarios: 
\begin{enumerate}
\item
The workplaces fully operational on November 1 and school/colleges open on January 1, 
\item
workplaces fully operational on November 1 and school/colleges remain closed, 
\item
workplaces and school/colleges fully operational from January 1. 
\end{enumerate}

Fully opening
workplaces or the economy implies that the  trains are back to the capacity as in normal pre-covid times. 
To err on the side of caution, we keep the train beta at a high risk level, that is, $\beta_T=0.4 \times \beta_H$, in 
these scenarios. 

\item
{\bf Containment zones and contact tracing and testing}  are regarded as two important policy tools available to decision makers in slowing the epidemic spread. Our small network framework (introduced  in \cite{City_Simulator_IISc_TIFR_2020}) through the  neighbourhood cells allows us to model neighbourhood containment efforts
with reasonable accuracy. Further, the small network framework with the introduction
of the community of friends and neighbourhood community, allows us to plausibly model the contact tracing and testing efforts. 
The methodology to aid in measuring contract tracing and testing in our model is introduced in \cite{City_Simulator_IISc_TIFR_2020}. This modelling feature is further discussed in Section~IV.
Through our model we evaluate the performance of containment efforts by 
measuring the health indicators as a function of varying containment efforts. 
We similarly evaluate the 
the impact of varying level of contact tracing and testing efforts
on the health indicators.    While containment zones are relatively easier to administer compared to contact tracing and testing, our simulations suggest that  the former may also be more 
effective in slowing the spread of the infection in the city. A caveat to keep in mind is that 
containment zones lead to restricting movements of relatively large number of people, and so  may
come at a significant economic and social cost.

\item
{\bf Impact of Vaccination:}  We consider the following simplistic scenarios:
Population above the age of 60 is vaccinated on February 1. The vaccine comes into effect
immediately and is 100\% effective so
that the vaccinated population instantly becomes  immune.
We also consider the scenario where population above the age of 50 is vaccinated
on February 1. Medical statistics under both these scenarios are compared 
to the statistics under the no vaccine setting.
\end{itemize}

\begin{figure}
  \begin{center}
    \small
    \begin{tabular}{|c|c|c|}
      \hline
      {\bf Age Group} & {\bf Slum age Distribution} & {\bf Non Slum Age Distribution}\\
      \hline
      Upto 10 yrs & 15.82\% & 15.82\%\\
      11-20 yrs & 18.38\% & 12.77\%\\
	  21-30 yrs & 18.51\% & 13.98\%\\
	  31-40 yrs & 17.70\% & 14.94\%\\
	  41-50 yrs & 11.35\% & 14.27\%\\
	  51-60 yrs & 11.35\% & 14.27\%\\
	  61-70 yrs & 5.48\% & 9.40\%\\
	  71-80 yrs & 1.17\% & 3.63\%\\
	  81 yrs and above & 0.21\% & 0.94\%\\
      \hline
    \end{tabular}
   \end{center}
  \caption{Age Distribution for Greater Mumbai. Age distribution for slums and non-slums 
  for 12 years and above is obtained from \cite{malani2020serosurvey}.
  We combine this  with \cite{mumbai_census2011_A} and \cite{suburban_mumbai_census2011_A}, assuming that both slums and non-slums
  have the same percentage of population younger than 12 years,  to arrive at the overall age distribution for Greater Mumbai.  }
  \label{Age-Distribution}
\end{figure}

\section{Small networks and containment}

\subsection{Smaller networks}

As detailed in our previous report, each of the interaction spaces is further broken down in subnetworks that corresponds to most of the
 interactions that an agent has within that interaction space (for instance, 90\% of an agent's interactions in the workplaces are within their ``project'' subnetwork). The subnetworks are listed as follows.

\begin{figure}
  \begin{center}
    \small
    \begin{tabular}{|ccc|}
      \hline
      {\bf Subnetwork} & {\bf Larger interaction space} & {\bf Description}\\
      \hline
      Project & Workplace & Clusters of size 3--10 (uniformly chosen)\\
      Class & School & All students of a specific age\\
      Neighbourhood & Community & Grid cells of side length 178 mts \\
      Close friends & Community & 2--5 households randomly chosen for each household\\
      \hline
    \end{tabular}
    
    \vspace{2em}
    
    \begin{tabular}{|ccc|}
      \hline
      {\bf Interaction space} & {\bf Comment}& {\bf $\beta$ value}\\
      \hline
      Home & (calibrated) & 1.93651 \\
      Workplace & (calibrated) & 0.26862\\
      Community & (calibrated) &  0.02152\\
      School & $2 \cdot \beta_{\text{workplace}}$ & 0.53723\\
      Project & $9 \cdot \beta_{\text{workplace}}$ & 2.41758\\
      Class & $9 \cdot \beta_{\text{school}}$ & 4.83507\\
      Neighbourhood & $9 \cdot \beta_{\text{community}}$ & 0.19368\\
      Close friends & $9 \cdot \beta_{\text{community}}$ & 0.19368\\
      \hline
    \end{tabular}
  \end{center}
  \caption{Interaction spaces, subnetworks and  contact rates  }
  \label{fig:beta-values}
\end{figure}

As in the earlier report~\cite{City_Simulator_IISc_TIFR_2020}, the contact rates are calibrated to match the observed growth of fatalities, and to have roughly equal contribution of infections from the household, community and workplace networks (including the subnetworks) in the ``no-intervention'' scenario. 

\subsection{Containment strategy}

While in  \cite{IISc_TIFR_2020_Mumbai_Report2} the containment zones were modelled at the ward level for computational ease, in 
the current implementation we aim for more accuracy through a finer and more accurate model of containment. 
In particular, we model containment at a neighbourhood cell level. Recall that our synthetic city
is divided into a grid of square cells where length of each cell is 178 meters. 
Containment effort  is modeled as an increasing adaptive  function of the active hospitalisations observed in the neighbourhood cell. Number of hospitalisations is taken as the decision variable since it is easily observable as compared to  tracking the number of positive cases in a cell, which may be harder to  estimate accurately without extensive testing, and the two are highly correlated. The cell
 is incrementally closed as more number of hospitalisations are observed in it.

Specifically, suppose that the   containment effectiveness (CE)  =75\% and there are 3,000 residents  in a neighbourhood cell. Then, 
first hospitalisation leads to movement restriction  of 25\% internally amongst the residents as well as to and fro from the cell; second hospitalisation leads 50\% movement restriction,
and third hospitalisation onwards leads to 75\% movement restriction. 
If, on the other hand, the neighbours in the cell are less than a thousand,  then as long as there exists a hospitalised person in the cell,
movement of every resident within  the cell, as well as movements  into and out of the cell are restricted by
 75\%.
 
More precisely,  If containment effectiveness is set to a fraction $y$, and
the neighbours in a cell equal $n$  thousand, then every hospitalisation  leads to $y/n$ restriction in movement, and total movement 
restriction
is capped at $y$.  Thus, 
percentage of activity restriction (internally as well as in entering and leaving the cell), or containment effectiveness, is our control and we set it  to
\[
\min\left( h y/n,  y\right),
\]
 where $h$ denotes the number of people in the neighbourhood cell  that are hospitalised\footnote{In our current implementation, the number of hospitalized cases \emph{excludes} those who are currently in critical care facilities.}.

\section{Simulation results}

As in \cite{IISc_TIFR_2020_Mumbai_Report2}, in all our simulations, outside of festival times, 
we set the compliance levels 
to 60\% in residential or non-slum
areas, and at 40\% in high density or slum areas.
This level of compliance with additional measures such
a mask usage, case-isolation and home quarantine
post lock-down, restriction on those above 65 to stay home, 
closed school and colleges match reasonably well
the observed data on fatalities.

Further, accounting for  the results of the Mumbai SeroSurvey~\cite{Mumbai_sero2020},
and in deviation from our earlier analysis in\cite{IISc_TIFR_2020_Mumbai_Report2},
we reduce the proportion of symptomatic population amongst those
exposed to the Covid-19 disease to 40\% from the earlier 66.67\%.
In addition,
in our earlier simulation runs
  for Mumbai in \cite{City_Simulator_IISc_TIFR_2020},  \cite{IISc_TIFR_2020_Mumbai_Report2},
  \cite{harsha2020covidmumbai} $\beta$ values for homes
  and communities in slums were kept at two times the values for non-slums. In the current simulations
  this factor is increased to 3. As mentioned earlier, this better matches the observed prevalance estimates 
  for Mumbai. The model is recalibrated to data using these adjustments.

\subsection{Long term projections}

We first discuss the long term projections. As mentioned earlier,
we consider these under the workplace attendance scenario
where after lockdown till May 17, there is 5\% attendance from May 18
to May 31st. This increases to 15\% attendance in June, 25\% in July,
33\% in August, 50\% in September and October and fully opens November onwards. 
In addition, as discussed earlier, we adjust for increase intermingling in population 
during the three festival times. 
These projections
are developed till March 15, 2021 under the following six scenarios.
\begin{itemize}
\item
The containment effectiveness kept at 75\% as well as 60\%.
\item
The infection rate from trains kept at base risk level $\beta_T=0.19 \times \beta_H$ (recall that  $\beta_H$ corresponds to the household transmission parameter). This was
derived as a plausible
rate of infection in trains in \cite{IISc_TIFR_2020_Mumbai_Report2}. To account for the uncertainty
in such calculations and population from outside Greater Mumbai using Mumbai locals, we also consider 
the more pessimistic medium risk 
setting of $\beta_T=0.30 \times \beta_H$ and high risk setting of $\beta_T=0.40 \times \beta_H$.
\end{itemize}

As in Reports \cite{City_Simulator_IISc_TIFR_2020} and \cite{IISc_TIFR_2020_Mumbai_Report2}, in our simulations, a synthetic city is created
that match the aggregate Mumbai demographic data. For this
city we run 5 independent simulations.
The reported results are the average of these five runs.

\begin{figure}
      \centering
      \begin{subfigure}[h]{0.7\linewidth} 
      \centering
 \includegraphics[width=\linewidth]{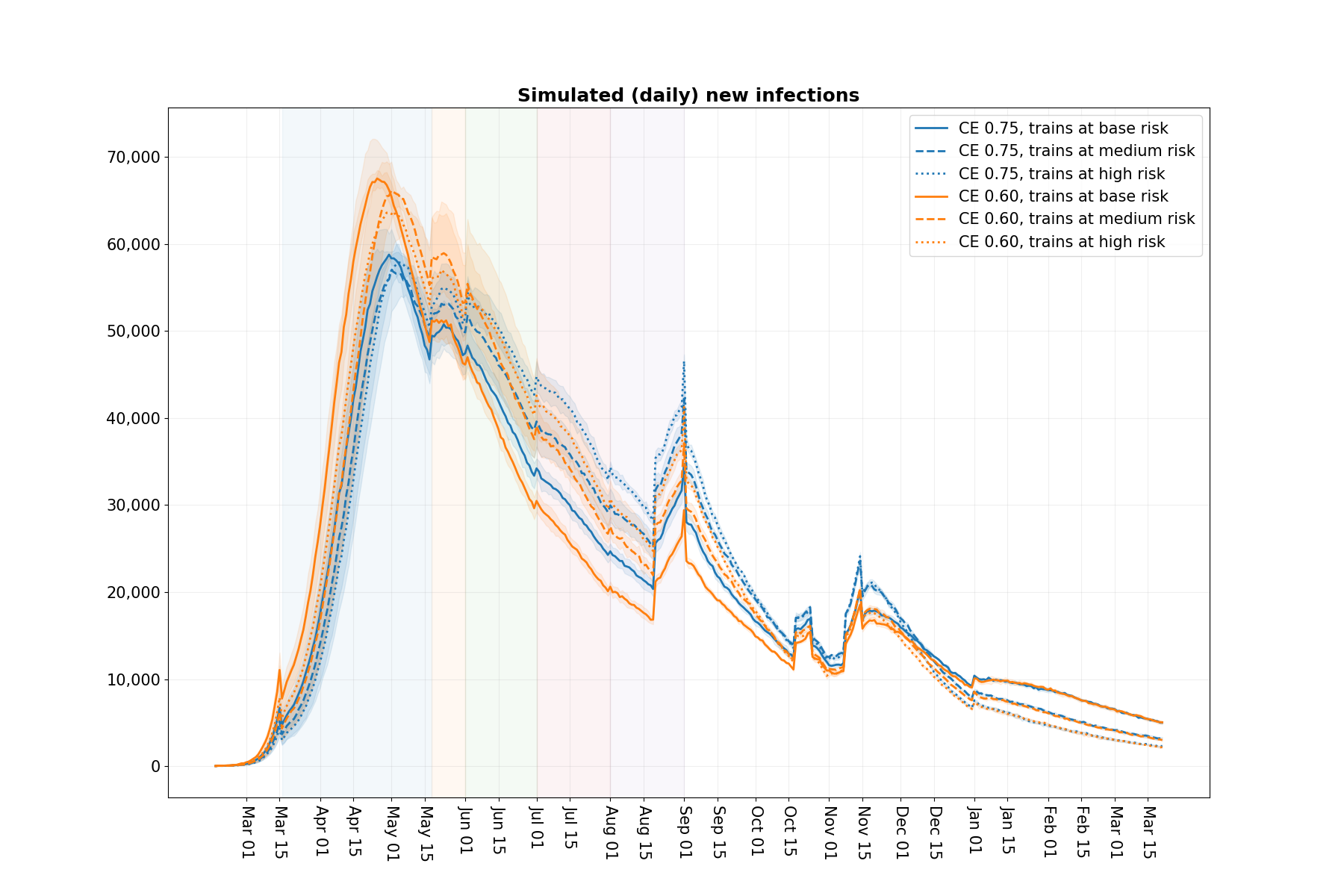}   
    \caption{Daily new infections
    under the workplace opening schedule 
     5\% attendance, May 18
to May 31st, 15\% attendance in June, 25\% in July,
33\% in August, 50\% in September and October and fully open November onwards. 
Includes Ganpati, Navratri/Dussehra and  Diwali relaxations.
    } \label{ltp_figure_daily_infections}
  \end{subfigure}

  \begin{subfigure}[h]{0.7\linewidth}
    \centering
    \includegraphics[width=\linewidth]{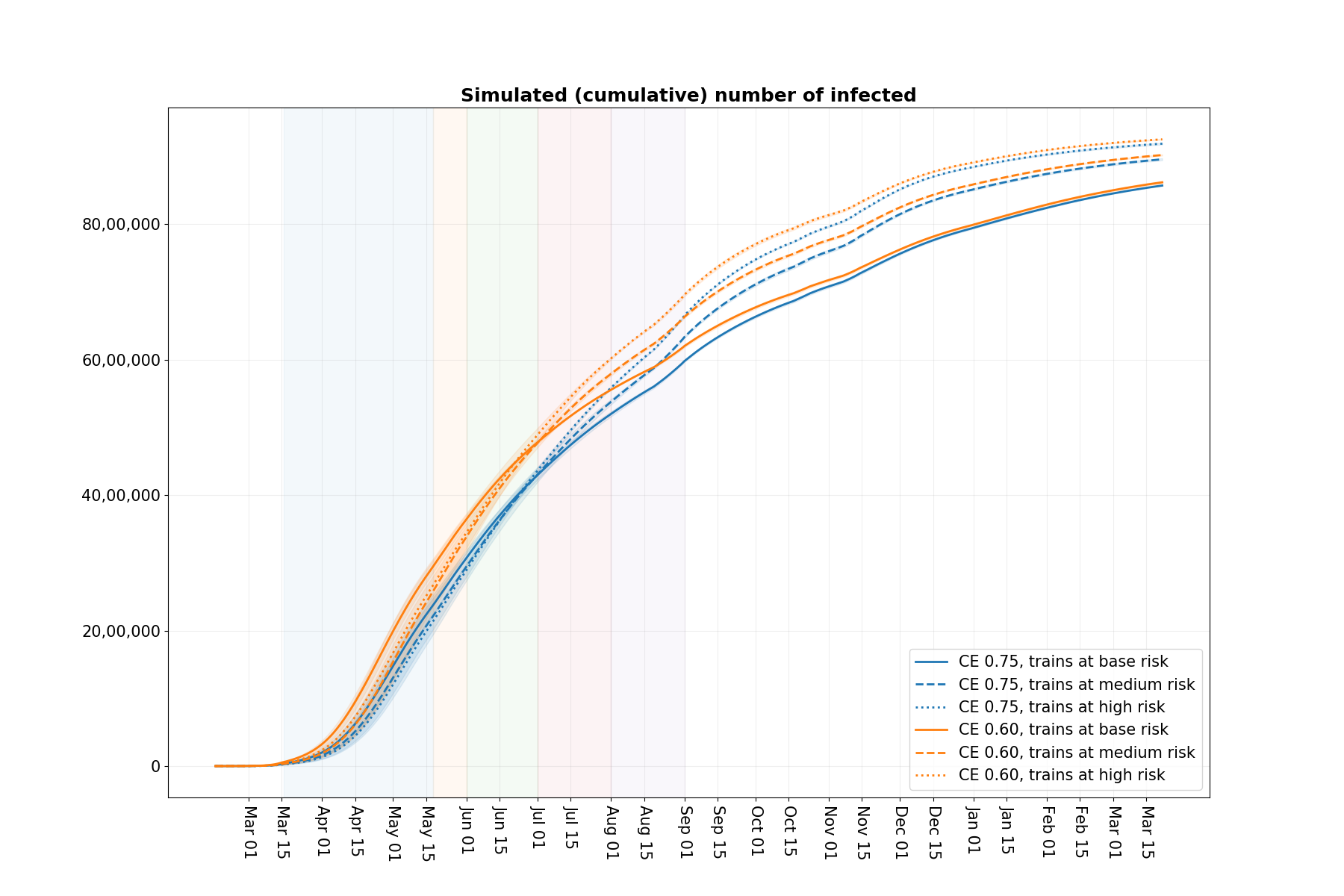}
   \caption{Cumulative  infection growth
    under the workplace opening schedule 
     5\% attendance, May 18
to May 31st, 15\% attendance in June, 25\% in July,
33\% in August, 50\% in September and October and fully open November onwards.
Includes Ganpati, Navratri/Dussehra and  Diwali relaxations.
Under this schedule
as per simulations the city stabilizes with 8 to 9 million of the population infected.}
\label{ltp_figure_cumulative_infections}
  \end{subfigure}
  \caption{}
  \end{figure}

In Figures~\ref{ltp_figure_daily_infections} and \ref{ltp_figure_cumulative_infections},
 we show the daily as well as the cumulative number of
infections under the six scenarios. These results suggest that
the growth of infections in Mumbai started to slow down from June, with 
slight increase every time the economy opened further or due to the festival season.
From January 1 onwards the number of new infections becomes very small
indicating that by and large herd immunity has been reached by the city.
Furthermore, this suggests that city will stabilize with close to 8-9 million residents infected.

\begin{figure}
      \centering
     \includegraphics[width=\linewidth]{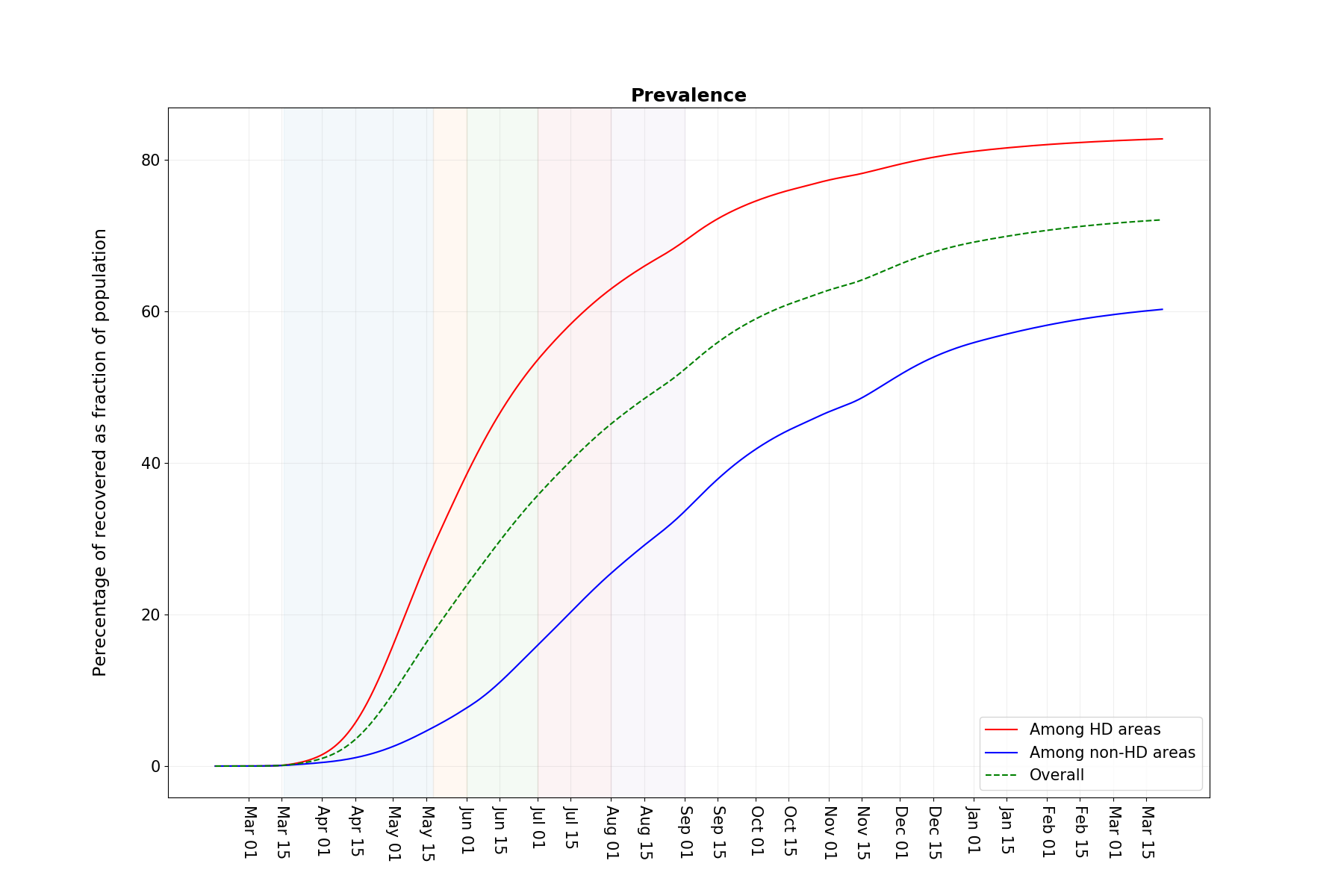}
      \caption{Simulated prevalence in Mumbai slums (HD areas) and non-slums
       under the workplace opening schedule 
     5\% attendance, May 18
to May 31st, 15\% attendance in June, 25\% in July,
33\% in August, 50\% in September and October and fully open November onwards. 
Includes Ganpati, Navratri/Dussehra and  Diwali relaxations.
 The herd immunity in slums 
is attained at around 80\%, while in non-slums
it is attained at prevalence close  to 55\%.} \label{ltp_figure_prevalence}
  \end{figure}

In Figure~\ref{ltp_figure_prevalence}, we map the prevalence for all of Mumbai suggested by the model when CE is set to 0.60 and $\beta_T$ is set at $0.4 \times \beta_H$.  We also separately
plot the prevalence for slums and non-slums.
The salient observations are that herd immunity is reached at different level
of prevalence in slums and non-slums. In slums this
is attained at around 85\%, while in non-slums
the number is closer to 60\%. The Mumbai SeroSurvey~\cite{Mumbai_sero2020} suggested 
around 55\% prevalence in slums and 15\% prevalence in
non-slums of the three wards of Mumbai that were sampled
around the first two weeks of July. Due to a typical gap of 1-2 weeks between the time the infection happens and
it can be detected by a serological test, we compare this with our model output on July 1. 
Our respective numbers on July 1 of 55\% and 15\% are more or less identical to  theirs.

\begin{figure}
    \begin{subfigure}[h]{\textwidth}
      \centering
     \includegraphics[width=\linewidth]{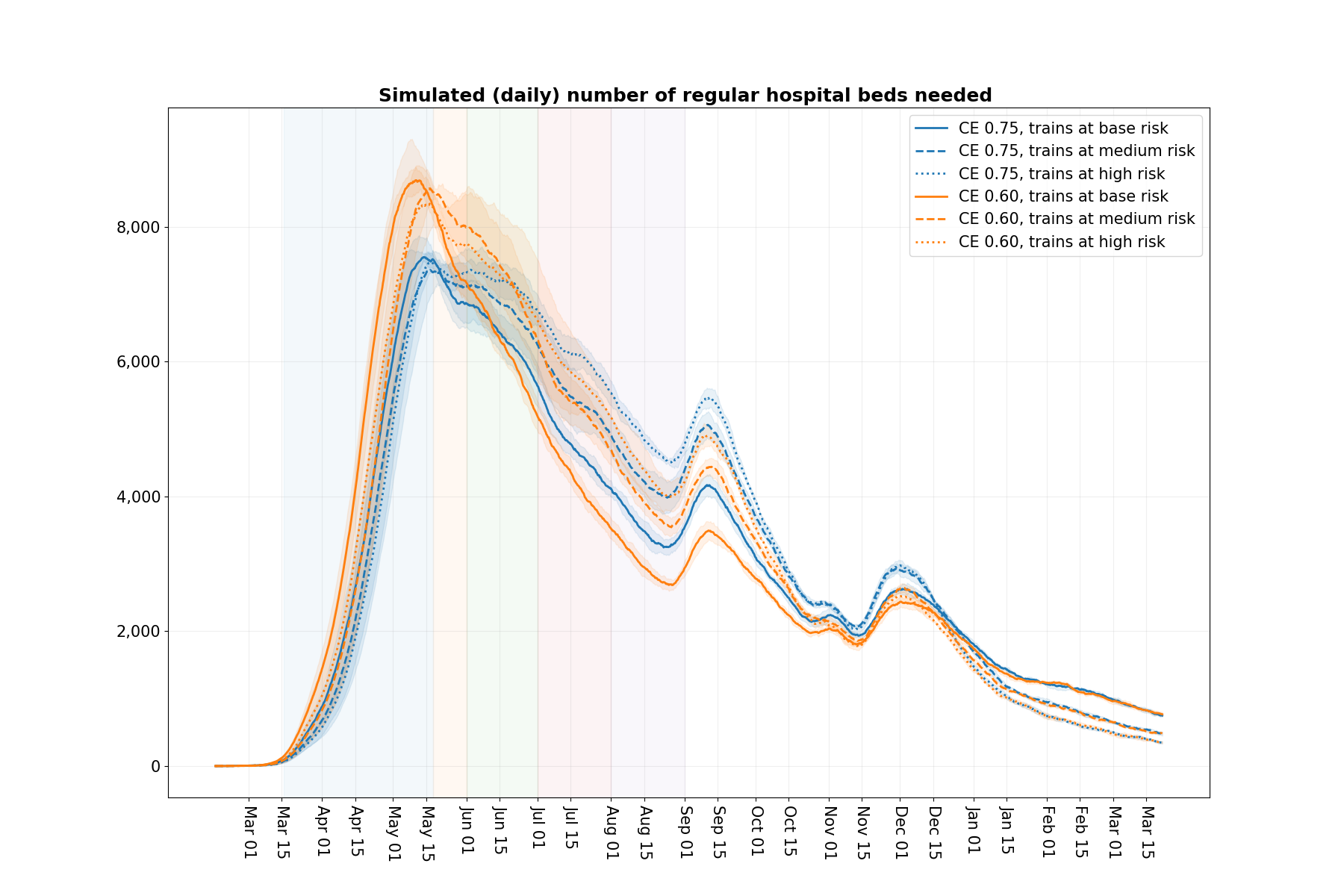}
     \caption{\footnotesize Simulated daily hospitalised patients in the city 
       under the workplace opening schedule 
     5\% attendance, May 18
to May 31st, 15\% attendance in June, 25\% in July,
33\% in August, 50\% till October and 100\% from Nov. 
Includes Ganpati, Navratri/Dussehra and  Diwali relaxations.}
     \label{ltp_figure_hospitalisation_a}
     \end{subfigure}%
     
\begin{subfigure}[h]{\textwidth}
      \centering
     \includegraphics[width=\linewidth]{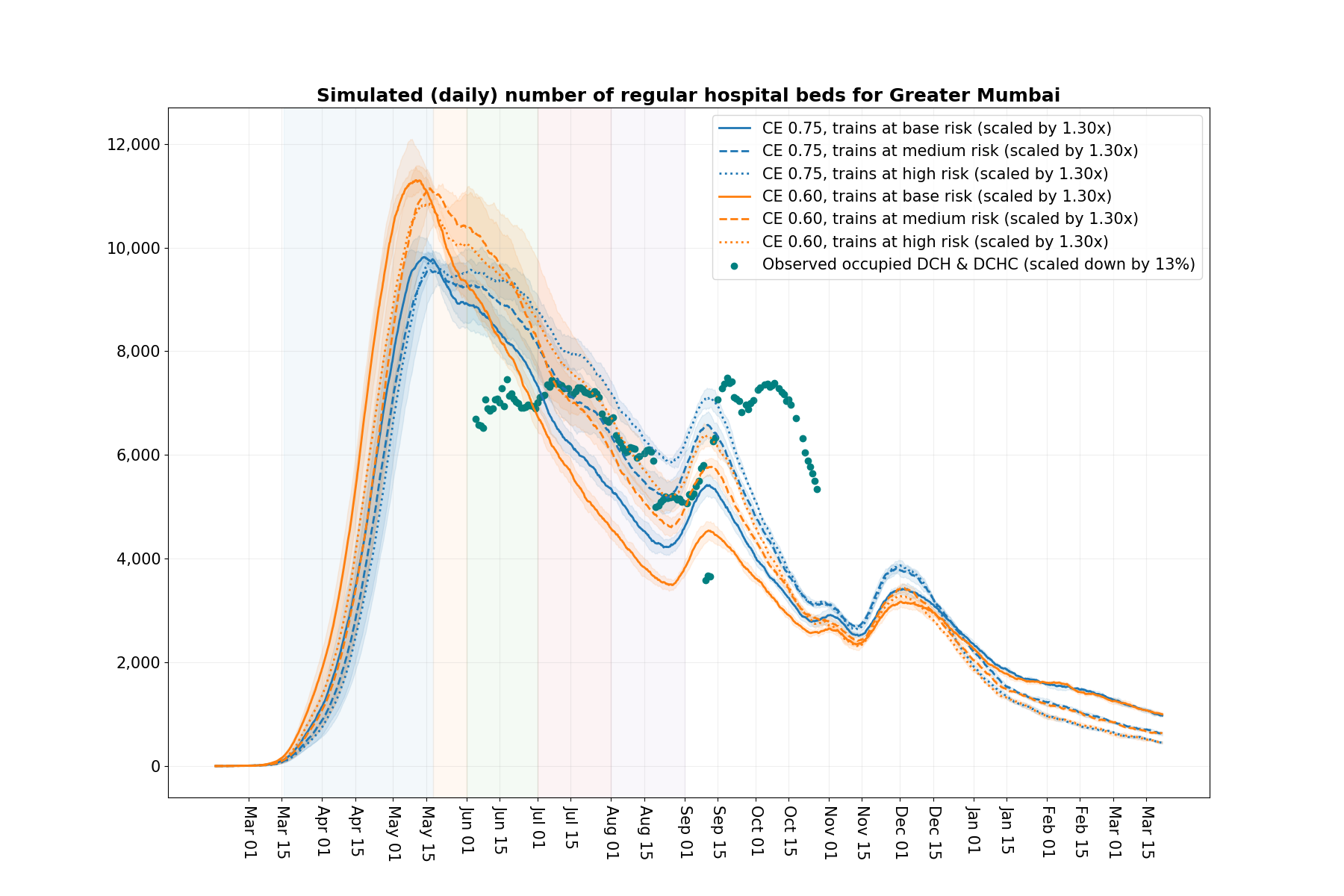}
      \caption{\footnotesize Comparison
      of simulated daily hospitalised patients in the city with the DCH and DCHC 
      numbers reported by BMC. The simulated numbers are increased by 30\% to account
      for estimated patients coming from the other MMR areas. These cases came to Mumbai
      mainly from around mid-July. The
      ICU numbers are removed from the reported DCH and DCHC numbers. These
       numbers are further reduced by 13\% to remove the estimated asymptomatic patients in DCH and DCHC to facilitate comparison.
      } \label{ltp_figure_hospitalisation_b}
    \end{subfigure}%
	\caption{}
  \end{figure}

Figure~\ref{ltp_figure_hospitalisation_a}  shows  the daily number of
the hospitalised patients  as per our model  under the six scenarios.
In Figure~\ref{ltp_figure_hospitalisation_b}, we compare
our hospitalisation numbers to the hospitalisation numbers
reported by BMC in their Dashboard \cite{BMCdashboard}. 
The BMC dashboard reports Dedicated Covid  Hospital (DCH) as well as Dedicated Covid Health Centre (DCHC) aggregated together and 
these are reported  as hospitalisations under DCH and DCHC.
We make the following adjustments to the data series  from our model as well as from the BMC Dashboard  
to make the comparison between them more apples to apples.
\begin{enumerate}
\item
As per personal communication with BMC, from
mid-July onwards many of the hospital beds are taken up
by patients coming from outside of Greater Mumbai (Greater Mumbai denotes the area that comes under the jurisdiction
of BMC). These 
include patients coming from other areas in
Mumbai Metropolitan Region  (MMR) including Thane, Navi Mumbai and
Vasai-Virar as well as from somewhat further regions such as Nasik. 
These are roughly estimated to equal 30\% (based on feedback from a BMC official). To account for these,
we increase our hospital patient projections by 30\%, with the understanding that this increase is
reasonable for comparison with observed data beyond mid-July.
\item
The DCH and DCHC numbers reported by the BMC Dashboard include patients under ICU. In our model we report
patients under ICU (critical patients) separately. Thus, we remove
the ICU patients in the Dashboard data from the DCH and DCHC data.
\item
Further, based on the snapshot data provided by BMC on August 1 and August
20, we inferred that about 13\%  of patients in DCH and DCHC are asymptomatic.
The disease progression data in our model is taken from \cite{ferguson2020report}
and \cite{verity2020estimates}, and here hospitalised patients correspond to those with serious symptoms.
 Thus, to compare our projections with the observed  data,
we further reduce the   DCH and DCHC  reported numbers by 13\%. 
\end{enumerate}

It can be seen from Figure~\ref{ltp_figure_hospitalisation_b} that with the above corrections,
post mid-July, our projections are reasonably close to the adjusted
DCH and DCHC numbers.

In Figure~\ref{ltp_figure_critical_a}, the projected daily number of
the critical cases   under the six scenarios are shown. Again, to compare 
with the Mumbai ICU numbers as per the BMC Dashboard, we scale our numbers 
by 41\% in Figure~\ref{ltp_figure_critical_b}. These again our based on a snapshot
input provided by BMC where 41\% of the critical beds used by the  Mumbai population were  in use by population from outside the city.
Here too, the match between the adjusted model and data after mid-July appears reasonable. From May 27 to 
June 16 
the occupancy of reported ICUs was seen to be above 98\% as per the BMC Dashboard (except on May 29, when  it was 97\%). This may at least partially explain why the model numbers for critical cases
in this period are much higher than the actual numbers.

Figure~\ref{ltp_figure_fatalities_a} shows the projected fatalities
based on our model and compares them
to the fatality data as reported by BMC through their Dashboard. 
Figure~\ref{ltp_figure_fatalities_b} shows cumulative number of fatalities
as a function of time. As pointed out in earlier reports \cite{City_Simulator_IISc_TIFR_2020} and \cite{IISc_TIFR_2020_Mumbai_Report2}, 
in our model the key transmission rates are calibrated to primarily match the 
initial observed fatality data in Mumbai as well as in the rest of India. Further, the compliance parameters used in the model
 are fine tuned
so that the model fatalities are close
to the reported fatality data (as reported by BMC through their Dashboard \cite{BMCdashboard}). As evident from the two figures, the match between the model generated fatality and the reported fatality data
appears to be quite good. Few points
are in order. 
\begin{itemize}
\item
BMC in mid-June had updated the reported deaths data. 
Figure~\ref{ltp_figure_fatalities_b} shows both the original and the updated reported fatality data. 
Observe,  that  our model (the data series corresponding to CE  0.60 and base or medium train risk  $\beta_T$ level) slightly underestimates
this series from mid-May to mid-June. One reason for this may be
that while in our model there is no limit on ICUs for critical patients,
the city of Mumbai did observe this shortage around that period.
\item
Our model reports higher number of deaths (under CE 60\%, and high train risk $\beta_T$ value) compared to the reported from July onwards. 
This may be partially explained by the fact that around mid-June
BMC stopped testing dead bodies  for covid \cite{TOI_article}.
\end{itemize}

A broad conclusion suggested by our model is that the fatalities
in the city will stabilize within 13,000 to 14,000 by March 2021.

In Figures~\ref{ltp_figure_cases_a} and ~\ref{ltp_figure_cases_b},
we report the observed daily  and the cumulative cases  in Mumbai. 
In Figure~\ref{ltp_figure_cases_a} we also report the daily tests conducted to illustrate
the correlation between these and the observed cases. This correlation
is especially strong from September first week to mid-October.
Since these numbers 
are a function of the testing strategy followed,
they are difficult to estimate. Later, in Section VI, we discuss
our contact  testing and tracing strategy that was tailored to 
the case data from mid-May onwards until the last ten days 
of August. However, since then the amount of testing in Mumbai has significantly 
increased and we do not have an accurate model to estimate the
number of positive cases thereafter.

In Figure~\ref{ltp_figure_slum_non_slum_fatalities} we report the time series of fatalities in slums and non-slums as per our 
model (the actual data is not available). The model result suggest that fatalities in slums
peaked during the months of May and June. They plateau around their peak from July to October
in non-slums.

{
  \begin{figure}
    \begin{subfigure}[h]{\textwidth}
      \centering
     \includegraphics[width=\linewidth]{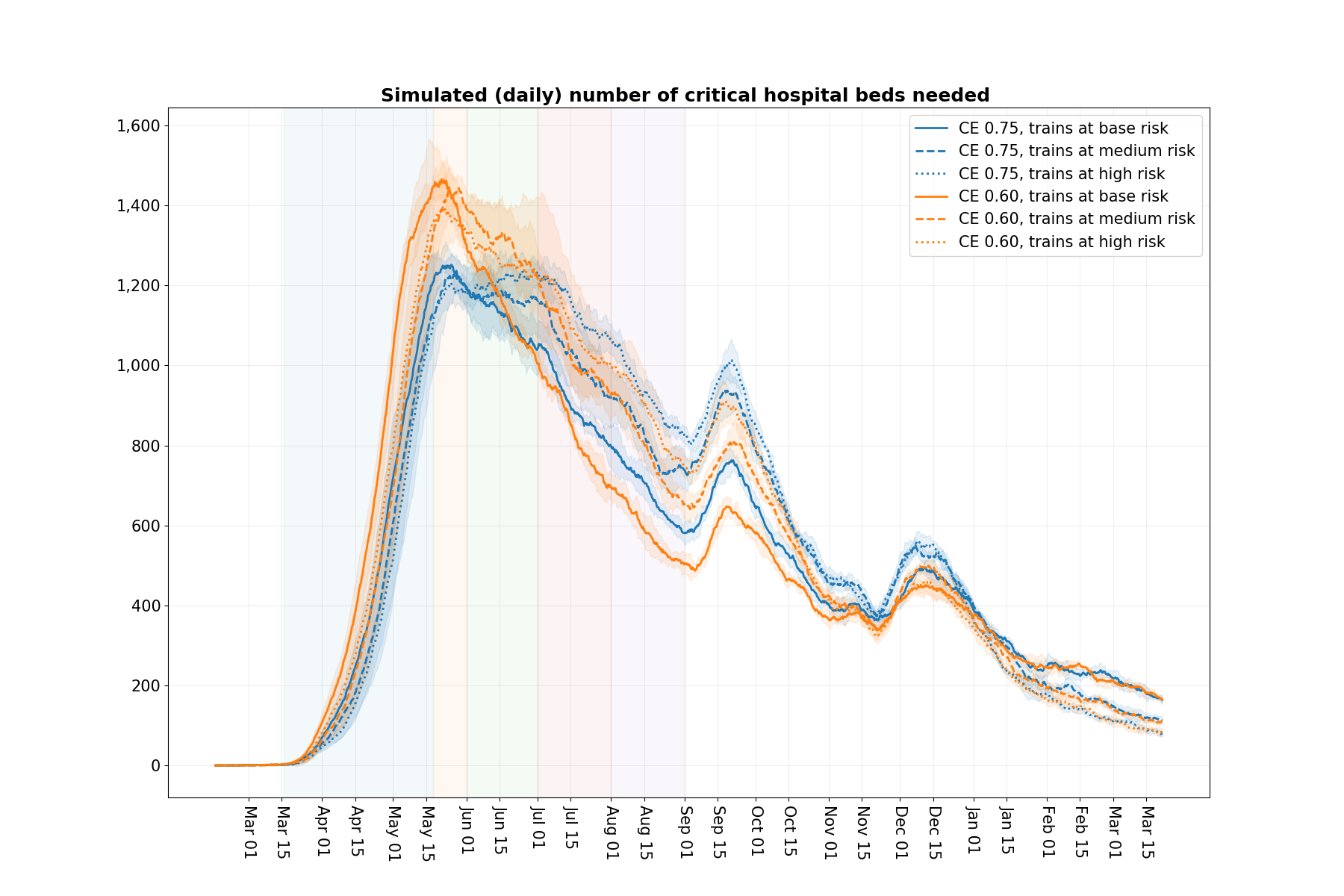}
      \caption{Simulated daily critical patients in the city 
       under the workplace opening schedule 
     5\% attendance, May 18
to May 31st, 15\% attendance in June, 25\% in July,
33\% in August, 50\% in September and October and fully open November onwards. 
Includes Ganpati, Navratri/Dussehra and  Diwali relaxations.}
     \label{ltp_figure_critical_a}
      \end{subfigure}%

\begin{subfigure}[h]{\textwidth}
      \centering
    \includegraphics[width=\linewidth]{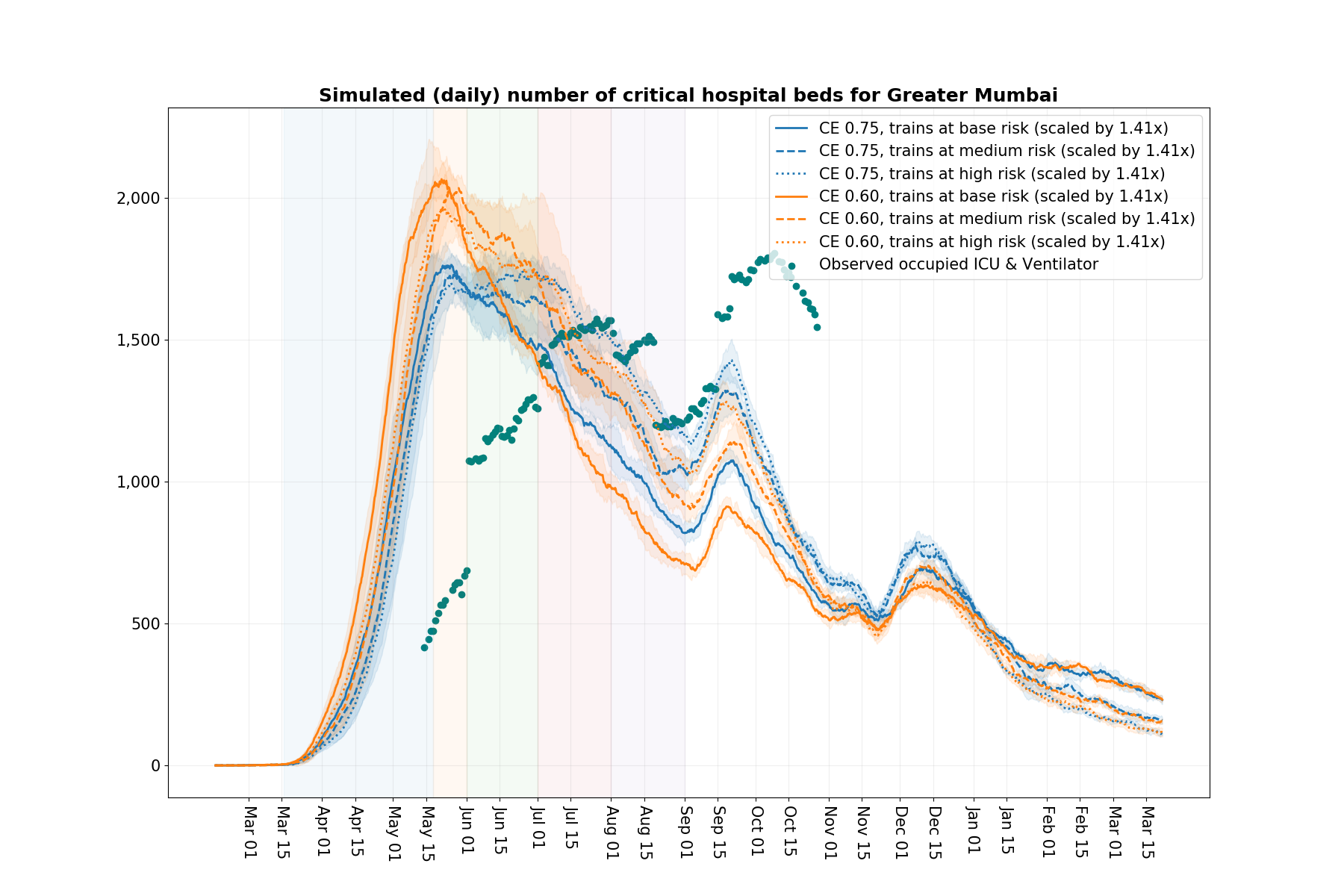}
    \caption{Comparison
      of simulated daily critical patients in the city with the DCH and DCHC 
      numbers reported by BMC. The simulated numbers increased by 41\% to account
      for estimated patients coming from the other MMR areas. These cases came to Mumbai
      from around mid-July. 
      } \label{ltp_figure_critical_b}
    \end{subfigure}%
    \caption{}
  \end{figure}
}

{
  \begin{figure}
    \begin{subfigure}[h]{\textwidth}
      \centering
         \includegraphics[width=\linewidth]{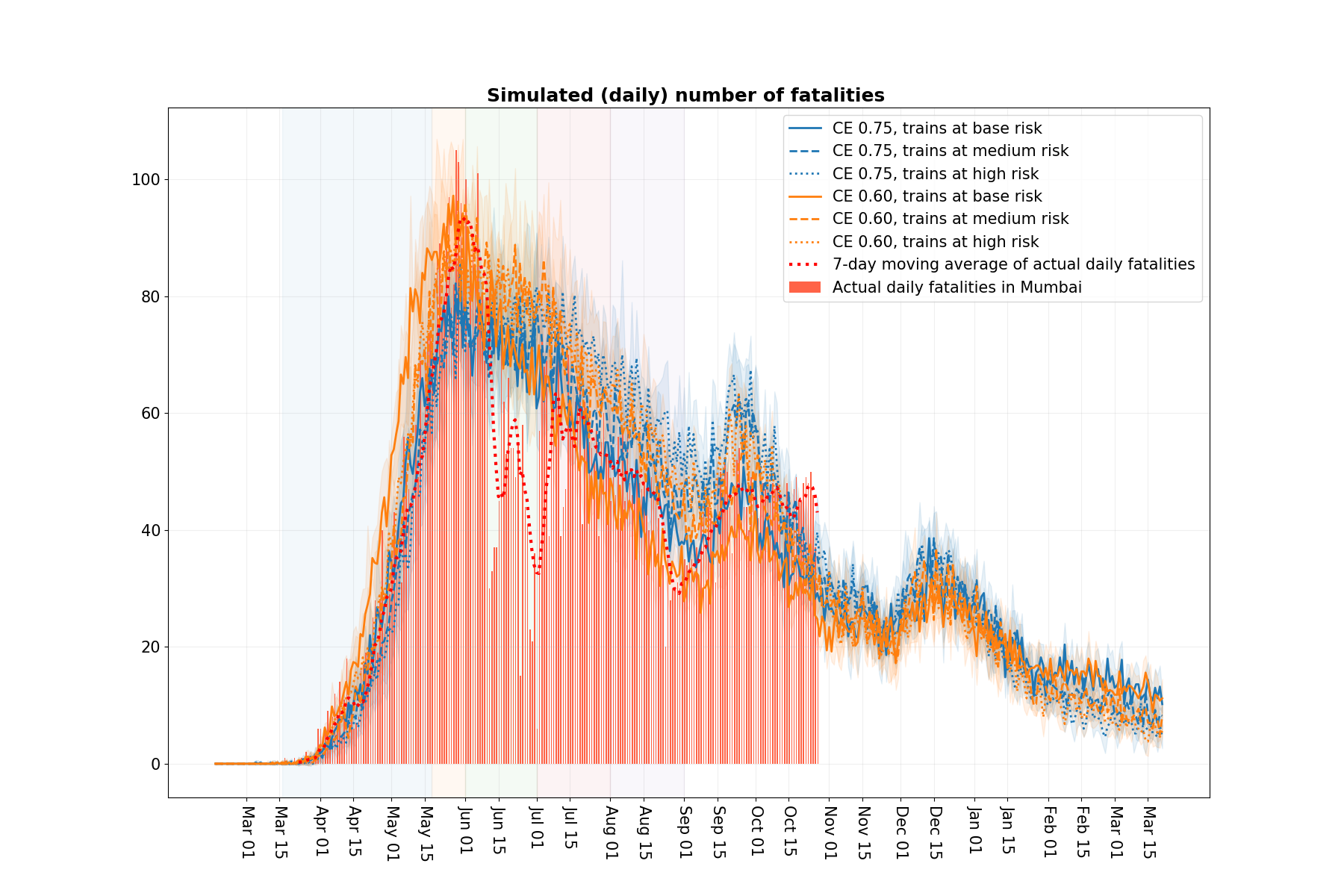} 
      \caption{Simulated daily deaths in the city 
       under the workplace opening schedule 
     5\% attendance, May 18
to May 31st, 15\% attendance in June, 25\% in July,
33\% in August, 50\% in September and October and fully open November onwards.
Includes Ganpati, Navratri/Dussehra and  Diwali relaxations.}
     \label{ltp_figure_fatalities_a}
      \end{subfigure}%
      
\begin{subfigure}[h]{\textwidth}
      \centering
      \includegraphics[width=\linewidth]{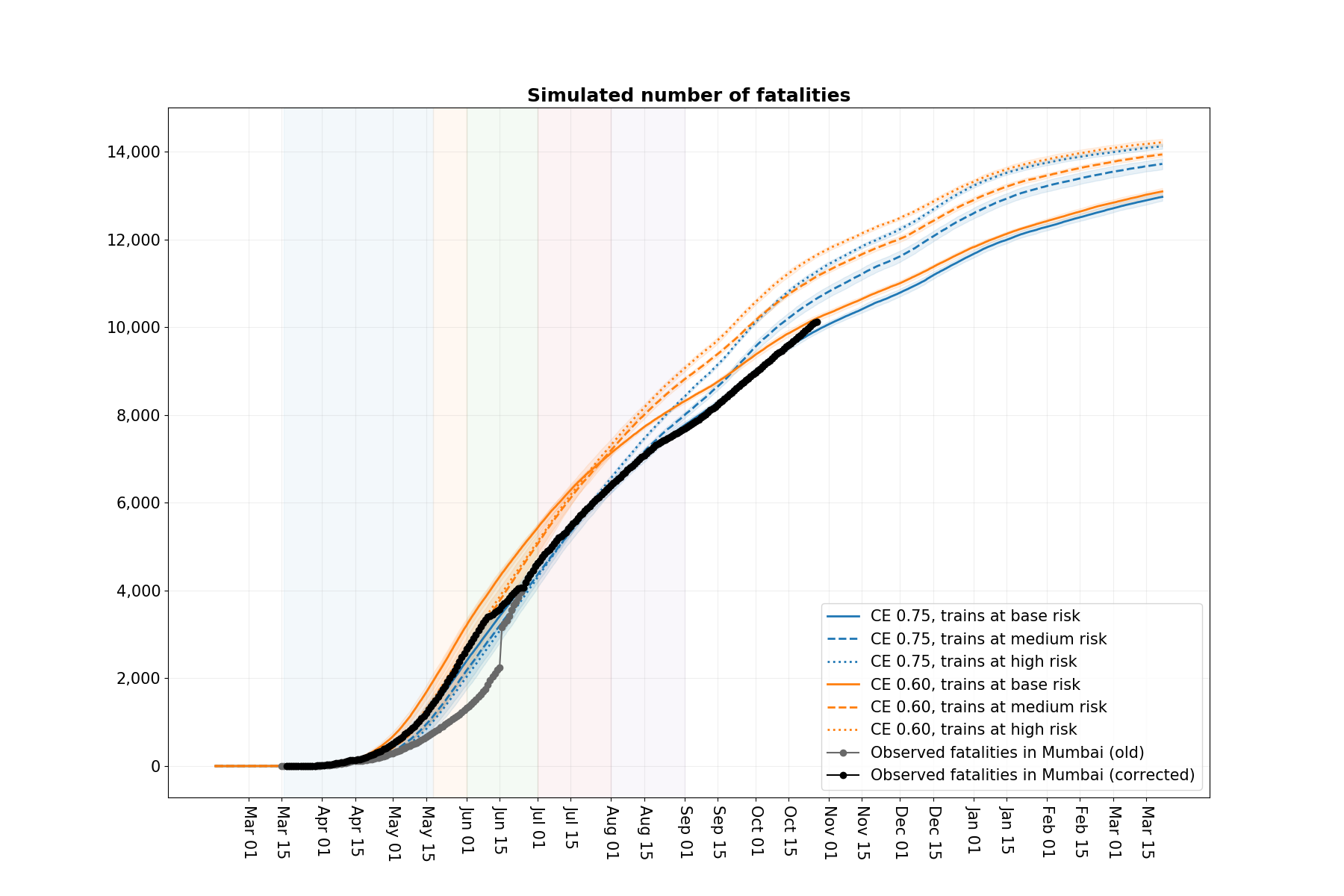}
      \caption{Simulated cumulative fatalities in the city 
       under the workplace opening schedule 
     5\% attendance, May 18
to May 31st, 15\% attendance in June, 25\% in July,
33\% in August, 50\% in September and October and fully open November onwards.
Includes Ganpati, Navratri/Dussehra and  Diwali relaxations. 
 As per the simulations, the fatalities in the city 
stabilize around 14,000 by March, 2021.
      } \label{ltp_figure_fatalities_b}
    \end{subfigure}%
    \caption{}
  \end{figure}
}

{
  \begin{figure}
    \begin{subfigure}[h]{\textwidth}
      \centering
         \includegraphics[width=\linewidth]{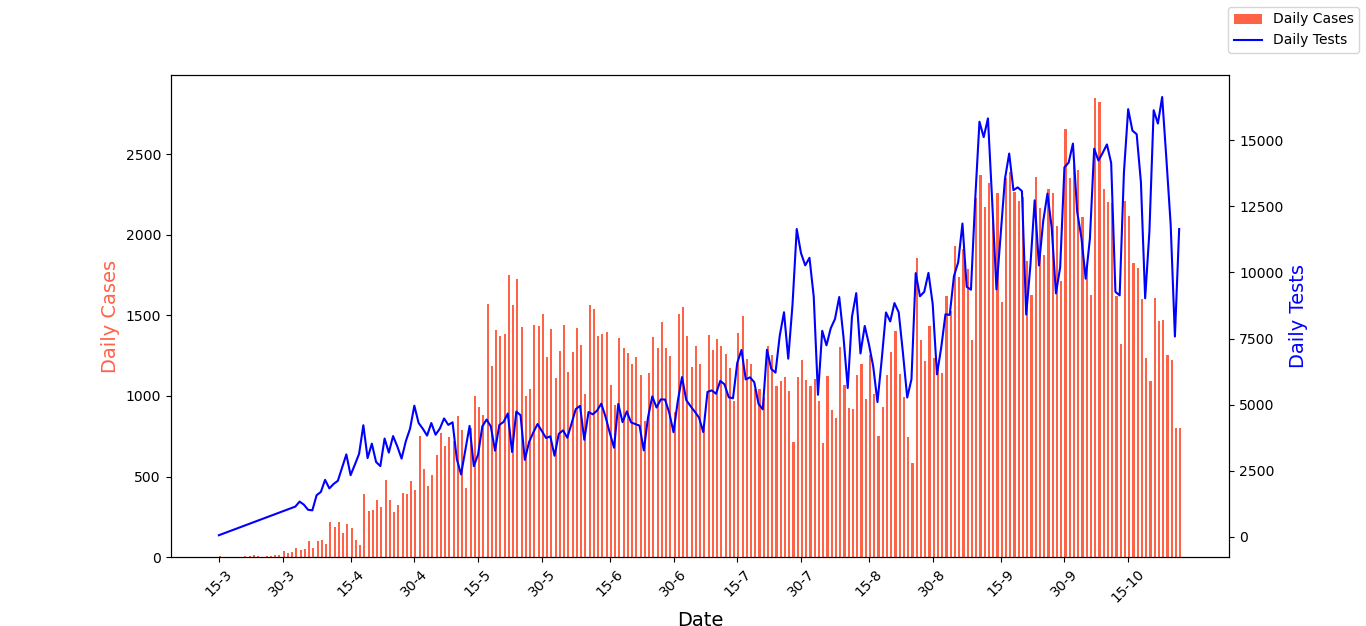} 
      \caption{Daily reported cases and tests in the city.  
        }
     \label{ltp_figure_cases_a}
      \end{subfigure}%
      
\begin{subfigure}[h]{\textwidth}
      \centering
      \includegraphics[width=\linewidth]{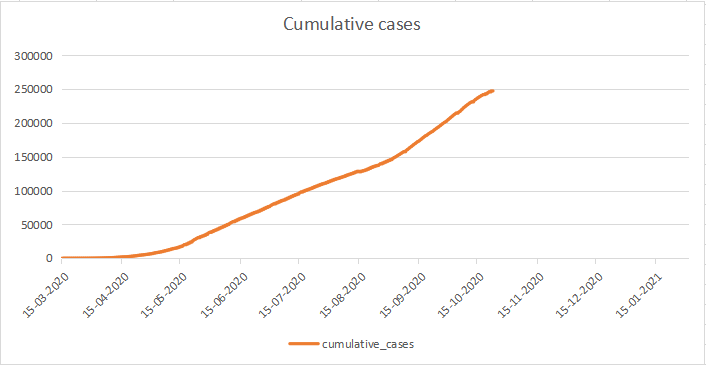}
      \caption{Cumulative reported cases in the city. 
        } \label{ltp_figure_cases_b}
    \end{subfigure}%
    \caption{}
  \end{figure}
}

\begin{figure}
      \centering
     \includegraphics[width=\linewidth]{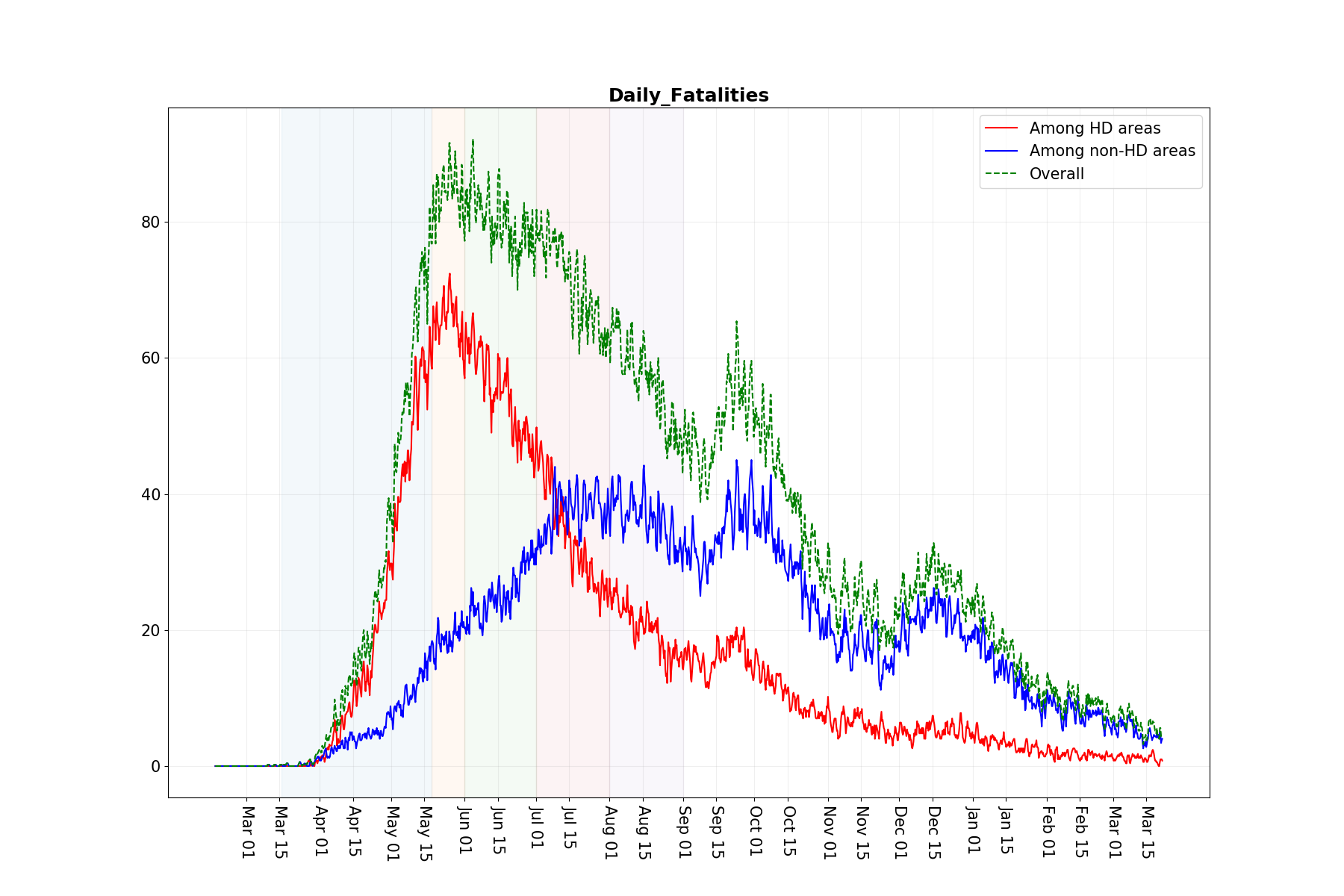}
      \caption{Simulated fatalities in Mumbai slums (HD areas) and non-slums
       under the workplace opening schedule 
     5\% attendance, May 18
to May 31st, 15\% attendance in June, 25\% in July,
33\% in August, 50\% in September and October and fully open November onwards.
Includes Ganpati, Navratri/Dussehra and  Diwali relaxations.
} \label{ltp_figure_slum_non_slum_fatalities}
  \end{figure}

\section{Fully operational of economic activity}

In Figure~\ref{open_figure_hospitalized}, we plot the simulated
hospitalisations  as well as critical cases where we compare the following three scenarios
\begin{itemize}
\item
Workplaces fully open on November 1 and School/Colleges open from January 1.
\item
Both Workplaces and School/Colleges fully open on January 1.
\item
Workplaces fully open on November 1 and School/Colleges remains closed.
\end{itemize}

All the scenarios include Ganpati, Navratri/Dussehra and  Diwali relaxations. 
 The simulated hospitalised numbers and critical numbers are increased by 30\% and 41\%, respectively, to account
      for estimated patients coming from the other MMR areas. The
      ICU numbers are removed from the reported DCH and DCHC numbers. These
       numbers are further reduced by 13\% to remove the estimated asymptomatic patients in DCH and DCHC.
       The train $\beta_T$ is set at high risk, that is $\beta_T= 0.4 \times \beta_{H}$.
        
Our key observations are that the second wave of hospitalisations and critical
cases is much higher with the November 1 opening compared
to the January 1 opening. While the projected hospitalisations
increase from around 2,300 a day to a peak of about 3,200  a day with the 
 November 1 opening, the increase is from about 200 a day to around 2,000 a day
 on January 1 opening. Further,  the opening of the schools on January 1 lead to only a small
increase in hospitalisations and critical cases.

Under the November 1 opening the daily critical cases peak at 
around 700 in mid-December under November 1 opening. These peak
at around 400 in mid-March after January 1 opening.

 Figure~\ref{open_figure_fatalities} reflects a similar pattern 
 in fatalities observed under these fully operational scenarios.
 While the projected daily fatalities
increase from around 20 a day to a peak of about 30 a day with the 
 November 1 opening, the increase is from about 4 a day to a peak of around 20 a day
 on January 1 opening. 

Figure~\ref{open_figure_exposed} show a similar pattern in number infected observed under these fully operational workplace scenarios.

{
  \begin{figure}
      \centering
     \includegraphics[width=\linewidth]{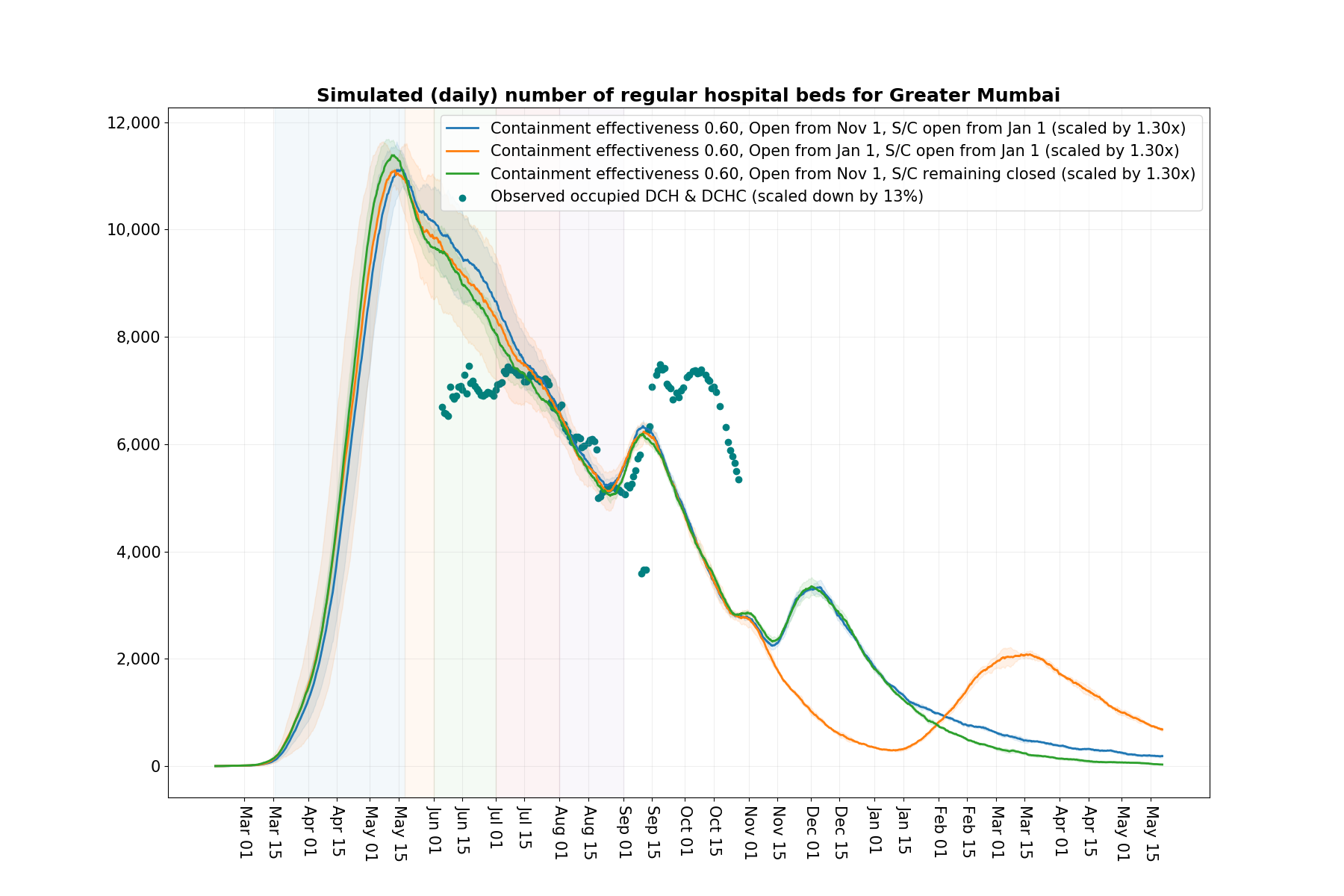}

    \includegraphics[width=\linewidth]{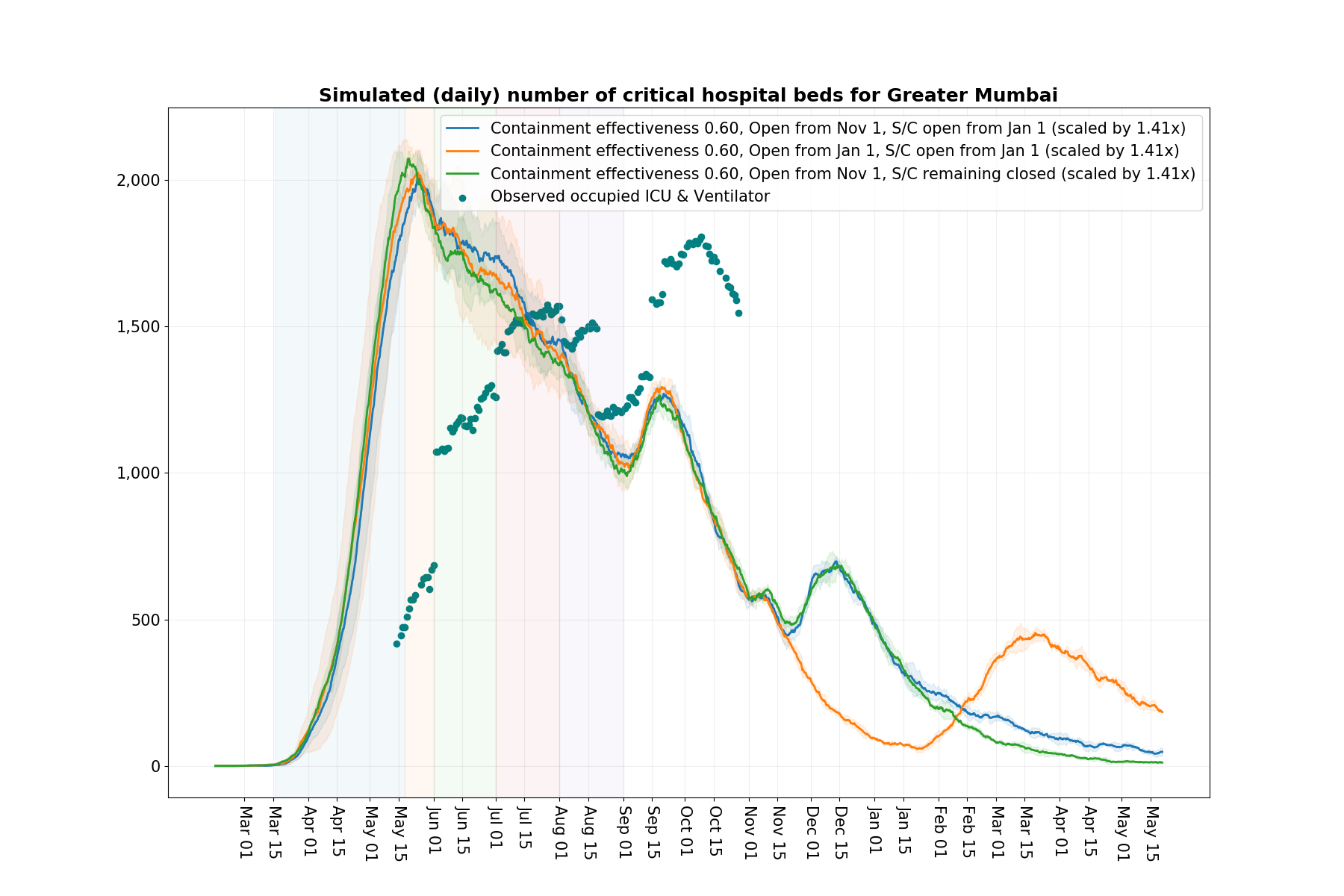}
      \caption{\footnotesize Simulated number of  daily hospitalized patients
      and daily critical cases under the workplace opening schedule 
     5\% attendance, May 18
to May 31st, 15\% attendance in June, 25\% in July,
33\% in August, 50\% in September and October and fully open November onwards with School/Colleges opening from January 1. This schedule
is overlaid with scenarios of 1) workplace and s/c opening from January 1, 2) Workplace open from November 1 and s/c remaining closed. All the scenarios include the three festival relaxations.  
 The simulated hospitalised numbers and critical numbers are increased by 30\% and 41\% respectively to account
      for estimated patients coming from other MMR areas. The
      ICU numbers are removed from the reported DCH and DCHC numbers. These
       numbers are further reduced by 13\% to remove the estimated asymptomatic patients in DCH and DCHC.  
      } \label{open_figure_hospitalized}
  \end{figure}
}

  \begin{figure}
      \centering
     \includegraphics[width=\linewidth]{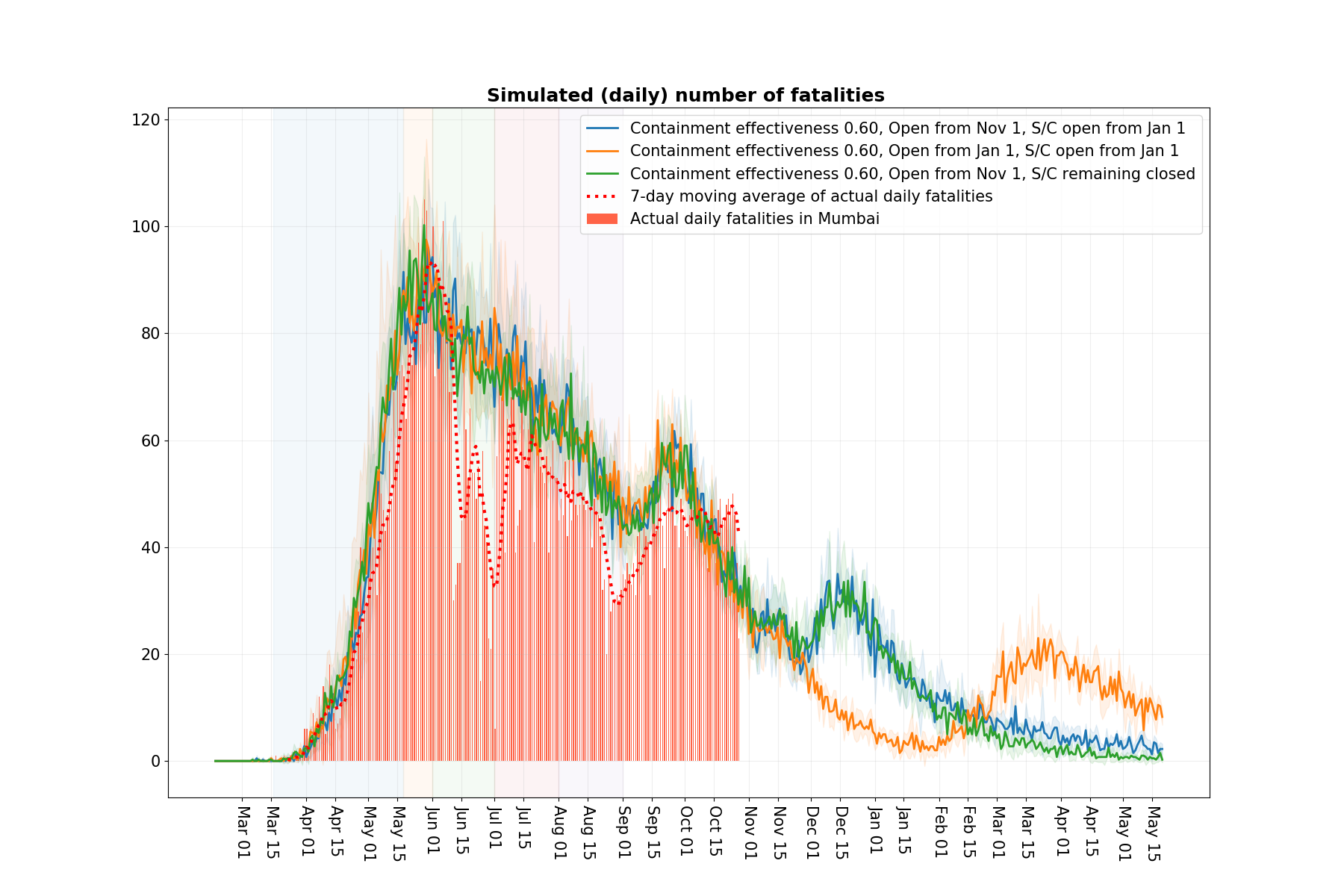}

    \includegraphics[width=\linewidth]{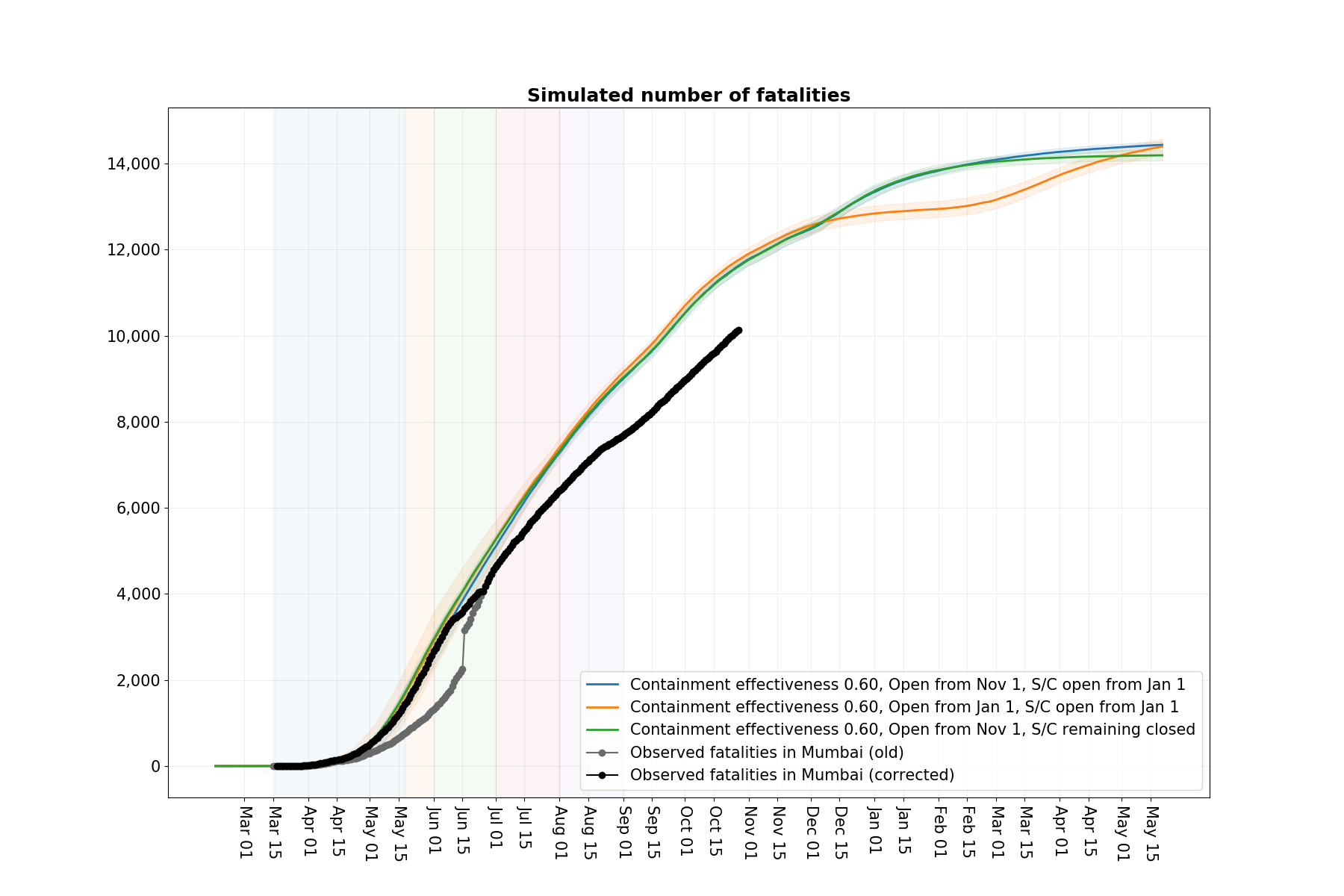}
      \caption{\small Simulated number of  daily and cumulative fatalities
 under the workplace opening schedule 
     5\% attendance, May 18
to May 31st, 15\% attendance in June, 25\% in July,
33\% in August, 50\% in September and October and fully open November onwards with School/Colleges opening from January 1. This schedule
is overlaid with scenarios of 1) workplace attendance of 100\% and school/colleges opening from January 1, 2) Workplace fully open from November 1 and school/colleges remaining closed. 
All the scenarios include the three festival relaxations.
} \label{open_figure_fatalities}
  \end{figure}

  \begin{figure}
      \centering
    \includegraphics[width=\linewidth]{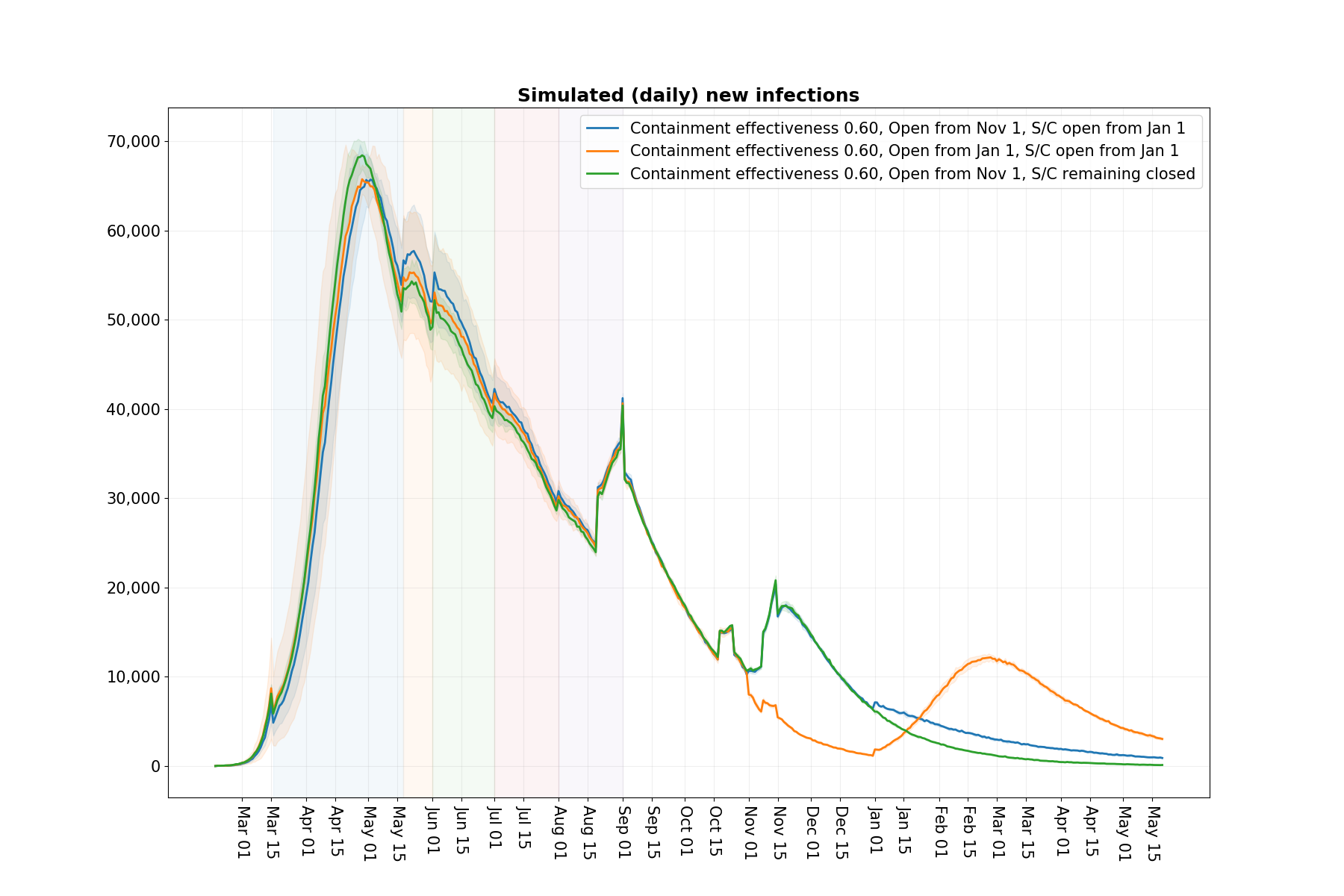}

\includegraphics[width=\linewidth]{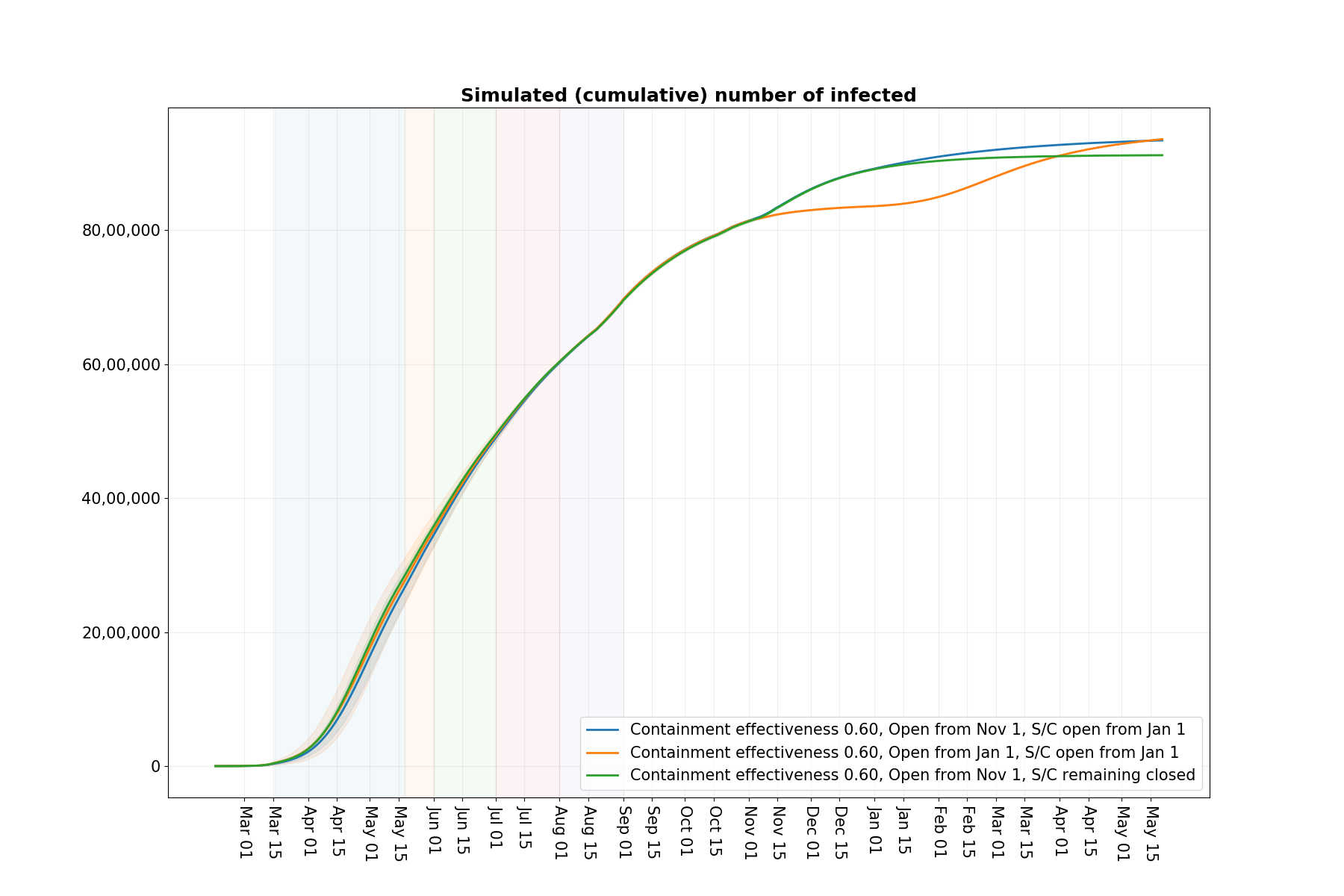}
      \caption{\small Simulated number of  daily and cumulative infections
 under the workplace opening schedule 
     5\% attendance, May 18
to May 31st, 15\% attendance in June, 25\% in July,
33\% in August, 50\% in September and October and fully open November onwards with School/Colleges opening from January 1. This schedule
is overlaid with scenarios of 1) workplace attendance of 100\% and school/colleges opening from January 1, 2) Workplace fully open from November 1 and school/colleges remaining closed. 
All the scenarios include the three festival relaxations.
} \label{open_figure_exposed}
  \end{figure}

\section{Comparing containment zones with contact tracing and testing}

For the comparison between the two strategies, containment zones are implemented as explained in Section~III, while contact tracing and testing strategy designed to reflect  the testing strategy and data  from mid-May to mid-August is explained below.

\subsection{Contact tracing and testing}

The contact tracing machinery can be briefly described as follows:
\begin{enumerate}\itemsep0pt
\item An individual that the simulator deems as a \emph{hospitalised} case undergoes a COVID-19 test with some probability (specified by the protocol given in
Figure~\ref{fig:testing-protocol}). 
If the test turns out to be positive, this agent is deemed as a \emph{hospitalised index case}.
This hospitalised case is typically tested with probability 1, although
later in Section  VI when we evaluate medical statistics under different testing protocols, we allow this probability
to take lower values of 0.66 and 0.8 as well.

\item When an \emph{index case} is identified, a fraction of agents from their subnetworks (specified by the protocol) are quarantined and marked as \emph{primary contacts}.
\item Each primary contact is tested with some probability (specified by the protocol). If the test is positive, then such agents are marked as \emph{positive index cases}. The newly discovered \emph{positive index cases} would additionally initiate contact trace around this agent like in the \emph{hospitalised index cases}.
\end{enumerate}

In the current implementation, 0.5\% of the neighbourhood cell (which is 5 agents on average) and 100\% of all other subnetworks are deemed as the agent's primary contacts. Furthermore, the testing probabilities in our current implementation are set to match a rough test positivity rate of 30\% to 40\%, which is the observed test positivity rate during the months of June and July in Mumbai \cite{BMCdashboard}.

\begin{figure}
  \begin{center}
    \small
    \begin{tabular}{|cccc|}
      \hline
      {\bf Subnetwork} & {\bf Type of index case} & {\bf Status of primary contact} & {\bf Test probability}\\
      \hline
      Household & Hospitalised & Symptomatic & 1 \\
                       & Hospitalised & Asymptomatic & 0.45\\
                       & Positive & Symptomatic & 1\\
                       & Positive & Asymptomatic & 0.45\\
      \hline
      Project & Hospitalised & Symptomatic & 0.5 \\
                       & Hospitalised & Asymptomatic & 0.225\\
                       & Positive & Symptomatic & 0.25\\
                       & Positive & Asymptomatic & 0.1125\\
      \hline
      Close friends & Hospitalised & Symptomatic & 0.25\\
                       & Hospitalised & Asymptomatic & 0.2\\
                       & Positive & Symptomatic & 0.125\\
                       & Positive & Asymptomatic & 0.06\\
      \hline
      Neighbourhood cell & Hospitalised & Symptomatic & 0.25\\
                       & Hospitalised & Asymptomatic & 0.2\\
                       & Positive & Symptomatic & 0.125\\
                       & Positive & Asymptomatic & 0.06\\
      \hline
    \end{tabular}
  \end{center}
  \caption{Testing protocol}
  \label{fig:testing-protocol}
\end{figure}

Figures~\ref{cz_figure_hospitalised} and \ref{cz_figure_fatalities}
show the impact of increasing containment effort on the resulting city medical statistics. Recall that containment efforts reduce the movement of individuals
within the neighbourhood containment cell as well as those going out from or coming into the containment cell. The graphs indicate that containment efforts go a long way in slowing the infection. Thus, containment  is an effective tool available to policy makers for slowing down the infection spread.

Figures~\ref{ctt_figure_hospitalised} and \ref{ctt_figure_fatalities}
show the impact of increasing contact tracing and testing  on the resulting city medical statistics. 
To test the intensity of contact tracing and testing, we consider three scenarios
where the hospital  index probability (the probability with which a hospitalised case is tested)
takes values 0.66, 0.80 and 1.0.

In these comparisons we have set the train transmission parameter $\beta_T$ to $0.3*\beta_H$. Further the relaxations due to the festivals are not considered.

The conclusion is that while contact tracing and testing does help in slowing the spread of the infection, the amount of reduction appears much less compared to that achieved through containment efforts, particularly 
since the latter appears to be cheaper and easier to implement.

The comparison between benefits of containment vis-a-vis
contact tracing and testing is crystallised
in Figure~\ref{figure-comparison} where we plot the peak of moving ten day average of daily hospitalised patients
 as a function of these efforts. The contact tracing and testing effort measured
on the x axis of the right hand figure corresponds to the hospital index
probability for values 0.66, 0.80 and 1.0. We also consider the two higher contact tracing and testing cases
where the hospital index
probability is kept fixed at 1 but the remaining testing probabilities in the protocol 
are increased by 25\% in one case and 50\% in the other.
These cases correspond to x-axis values of 1.25 and 1.5 in Figure~\ref{figure-comparison}.

{
  \begin{figure}
      \centering
     \includegraphics[width=\linewidth]{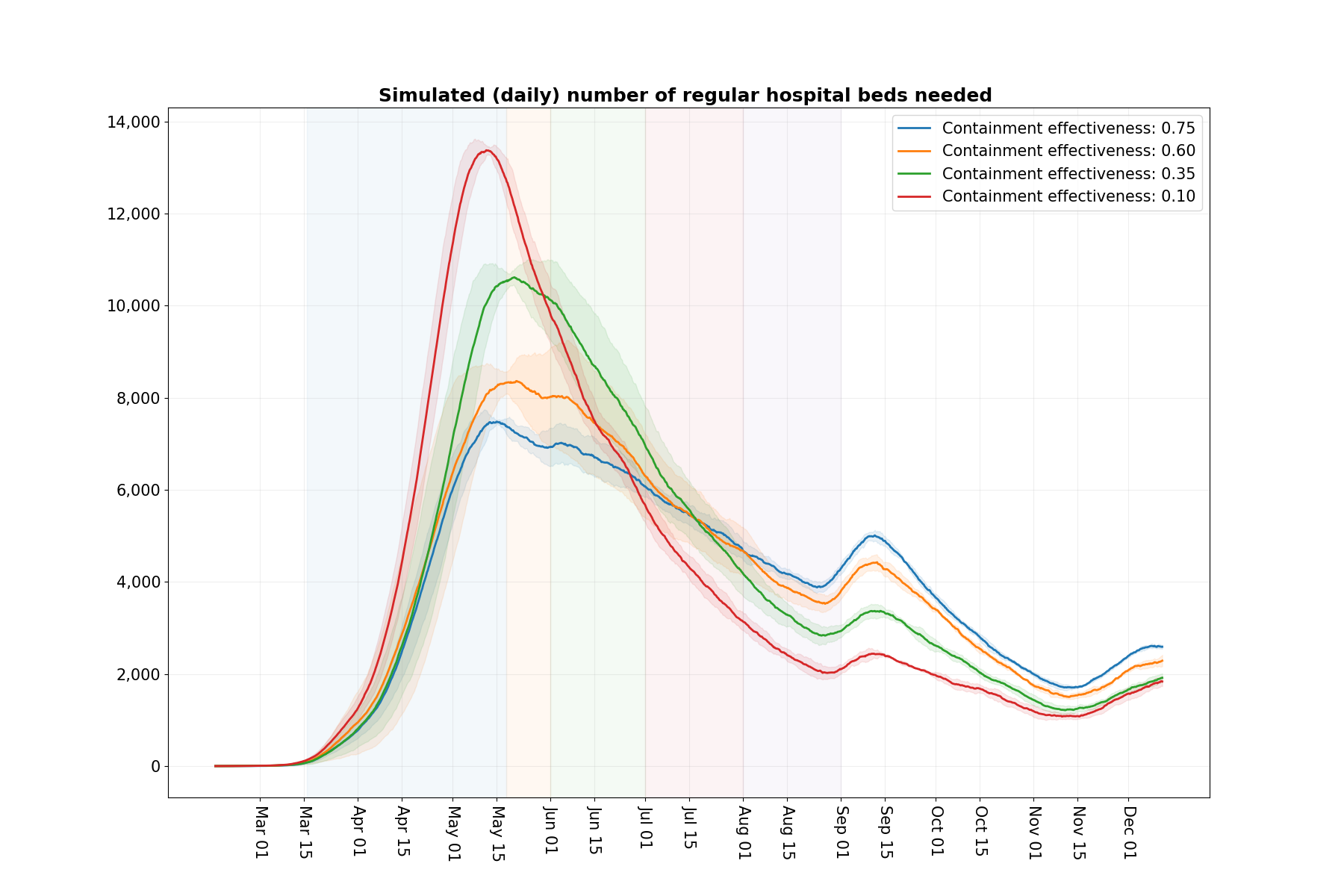}
    \includegraphics[width=\linewidth]{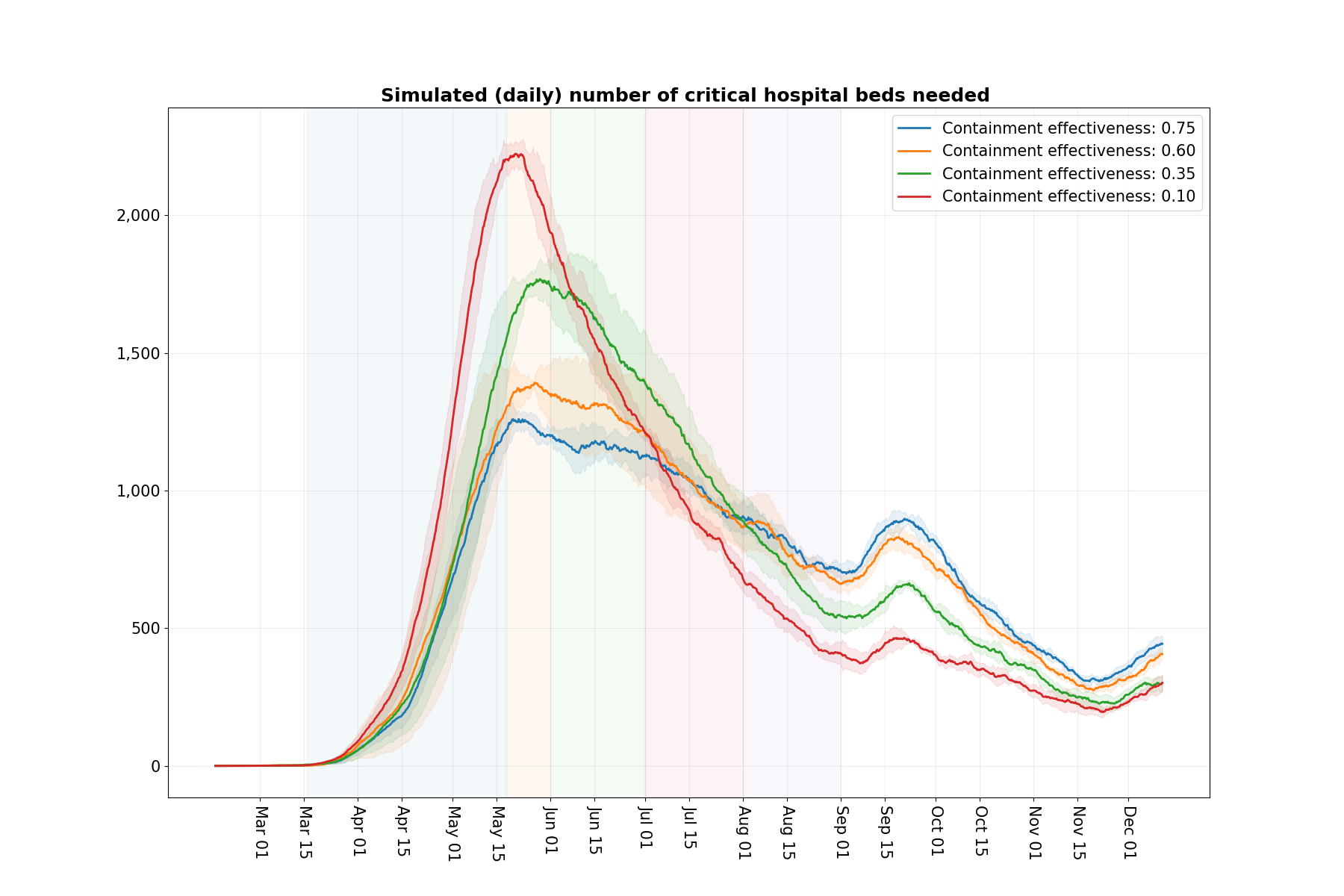}
      \caption{Simulated number of  daily hospitalised patients and critical patients
under varying level of containment efforts.       
 Workplace opening schedule 
     is set at 5\% attendance, May 18
to May 31st, 15\% attendance in June, 25\% in July,
33\% in August, 50\% in September and October and fully open November onwards.
} \label{cz_figure_hospitalised}
 \end{figure}
 }

{
  \begin{figure}
      \centering
       \includegraphics[width=\linewidth]{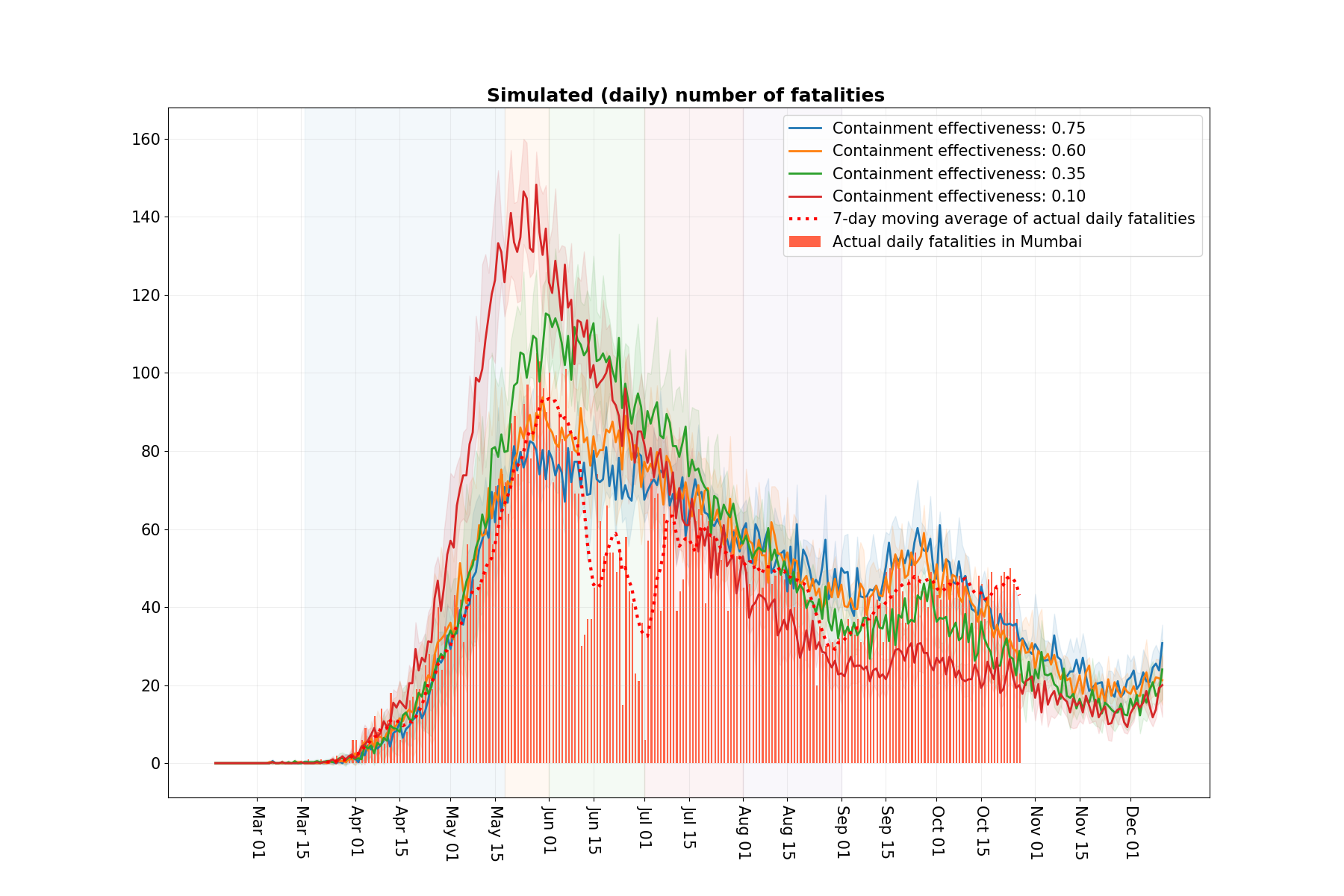}
      \caption{Simulated number of  daily fatalities under varying level of containment efforts.       
 Workplace opening schedule 
     is set at 5\% attendance, May 18
to May 31st, 15\% attendance in June, 25\% in July,
33\% in August, 50\% in September and October and fully open November onwards.  
 } \label{cz_figure_fatalities}
  \end{figure}
}

{
  \begin{figure}
      \centering
     \includegraphics[width=\linewidth]{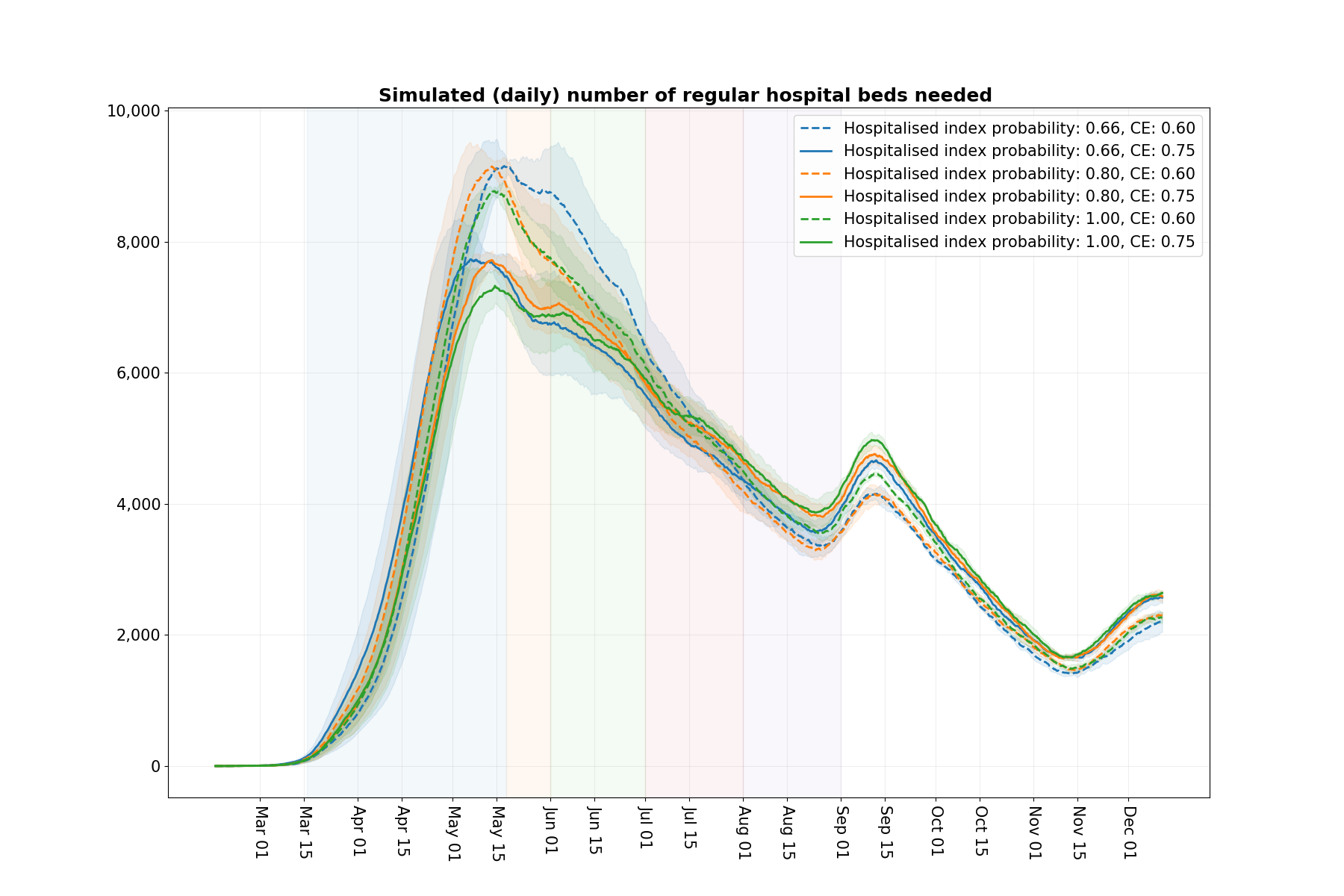}
    \includegraphics[width=\linewidth]{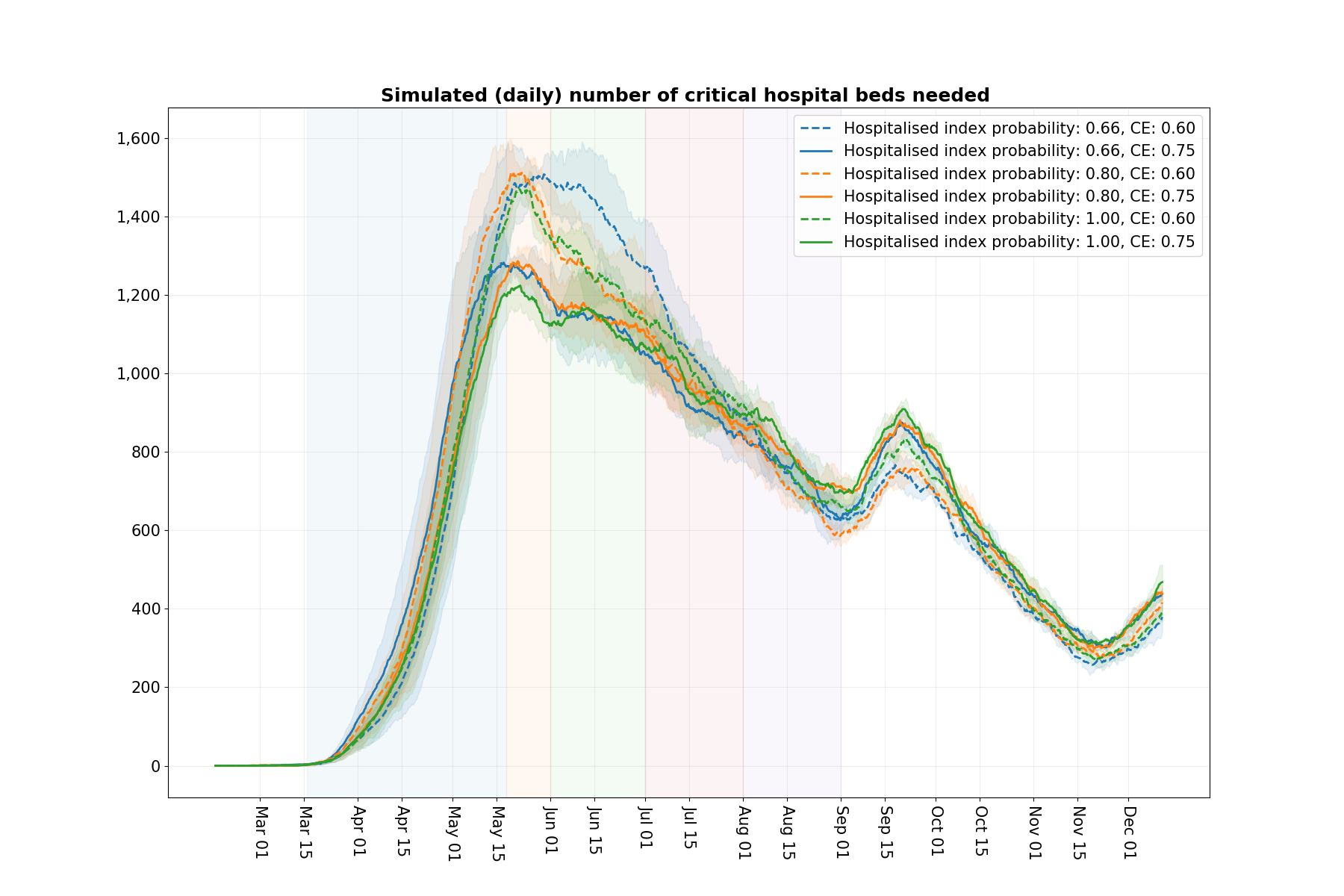}
      \caption{Simulated number of  daily hospitalised patients and critical patients
under varying level of contact tracing and testing strategies.       
 Workplace opening schedule 
     is set at 5\% attendance, May 18
to May 31st, 15\% attendance in June, 25\% in July,
33\% in August, 50\% in September and October and fully open November onwards. } \label{ctt_figure_hospitalised}
    \end{figure}
} 

{
  \begin{figure}
      \centering
       \includegraphics[width=\linewidth]{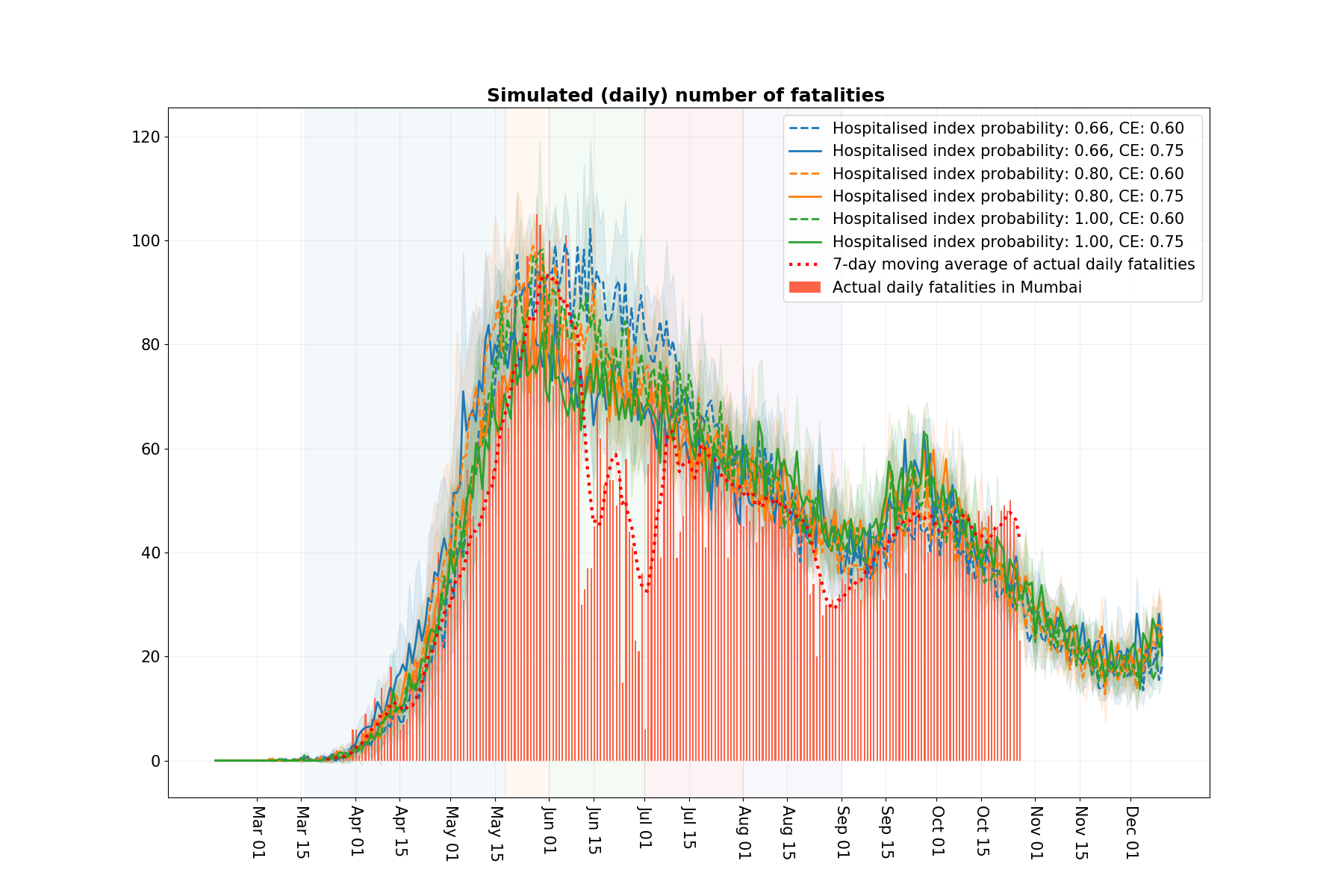}
      \caption{Simulated number of  daily fatalities under varying level of contact tracing and testing strategies.       
 Workplace opening schedule 
     is set at 5\% attendance, May 18
to May 31st, 15\% attendance in June, 25\% in July,
33\% in August, 50\% in September and October and fully open November onwards.} \label{ctt_figure_fatalities}
  \end{figure}
}

{
  \begin{figure}
      \centering
       \includegraphics[width=\linewidth]{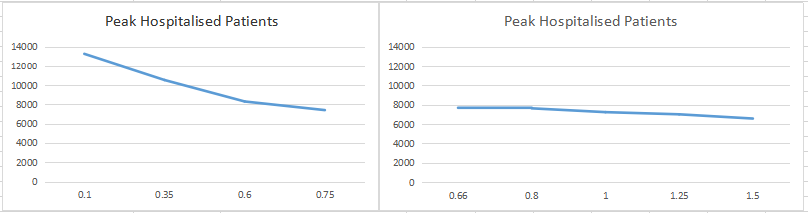}
      \caption{Comparison between benefits of containment effort vis-a-vis
contact tracing and testing. The left figure reports the
 peak of moving ten day average of daily hospitalised patients
 as a function of containment effort. The right figure reports the same peak as a function
 of contact tracing and testing efforts.} \label{figure-comparison}
  \end{figure}
}

\section{Impact of Vaccination}

Figures~\ref{vaccination_figure_hospitalised} and \ref{vaccination_figure_fatalities}
show the impact of introducing vaccination on the resulting city medical statistics. 
The three festival relaxations are included in these simulations.
The train transmission parameter $\beta_T$ is set to its medium value $0.3*\beta_H$.
As mentioned earlier, to keep the discussion simple,
we assume vaccines are administered to the specified population on February 1.
They work instantly and provide complete immunity at least for the next six months.

Figure~\ref{fatalities_vaccination} suggests that by vaccinating people aged 60 years and above on Feb 1, 2021, around 498 lives (appx. 53\%) can be saved in the next six months whereas vaccinating people aged 50 years above saves around 607 lives (appx. 64\%) over the same period.  
It can also be seen from the simulation results that introducing vaccination significantly reduces the load on medical 
facilities. By vaccinating people aged 60 years and above on Feb 1, 2021, the hospitalisations (including critical
cases) reduces to 5342 (from 8840 in the no vaccination case) which amounts to appx. 39.6\% reduction. Similarly, by vaccinating people aged 50 years and above, the  hospitalisations (including critical cases) reduces to 2908 (appx. 67\% reduction).  

The demographics of Greater Mumbai as estimated from Figure 1 suggests that 
approximately 13.1 lac Mumbai residents are aged 60 years or above. 
 According to our simulations, out of these 13.1 lac, approximately 5.6 lac will be susceptible on Feb 1. Similarly, the total number of residents aged 50 and above  is 29.4 lac whereas around 10.3 lac will be susceptible on Feb 1. Since, we may not be able to, or it may not even be desirable to, separate the susceptibles
from the recovered population from vaccination viewpoint,  
we may have to vaccinate everyone in the respective age group to provide them immunity.

{
  \begin{figure}
      \centering
     \includegraphics[width=\linewidth]{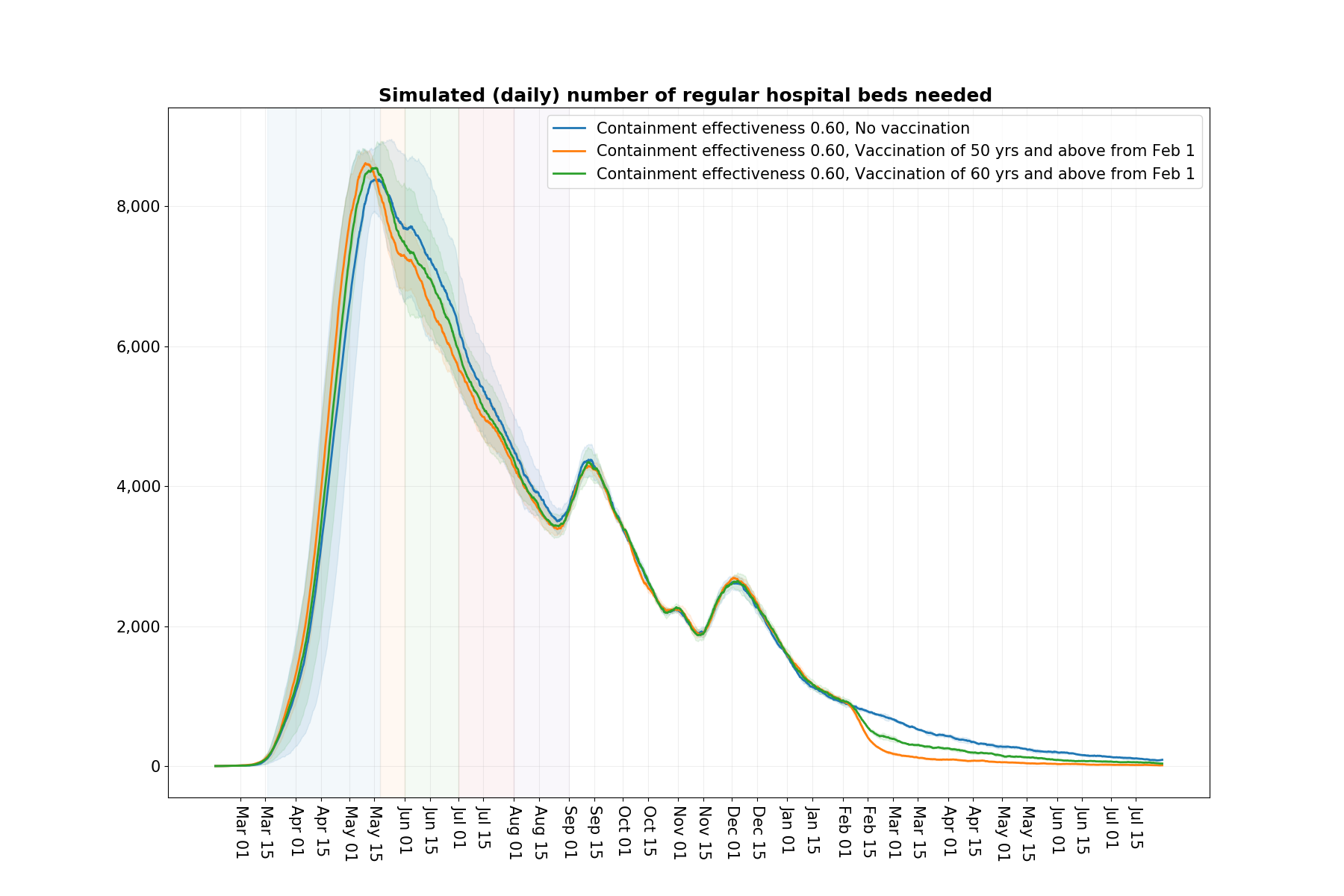}
    \includegraphics[width=\linewidth]{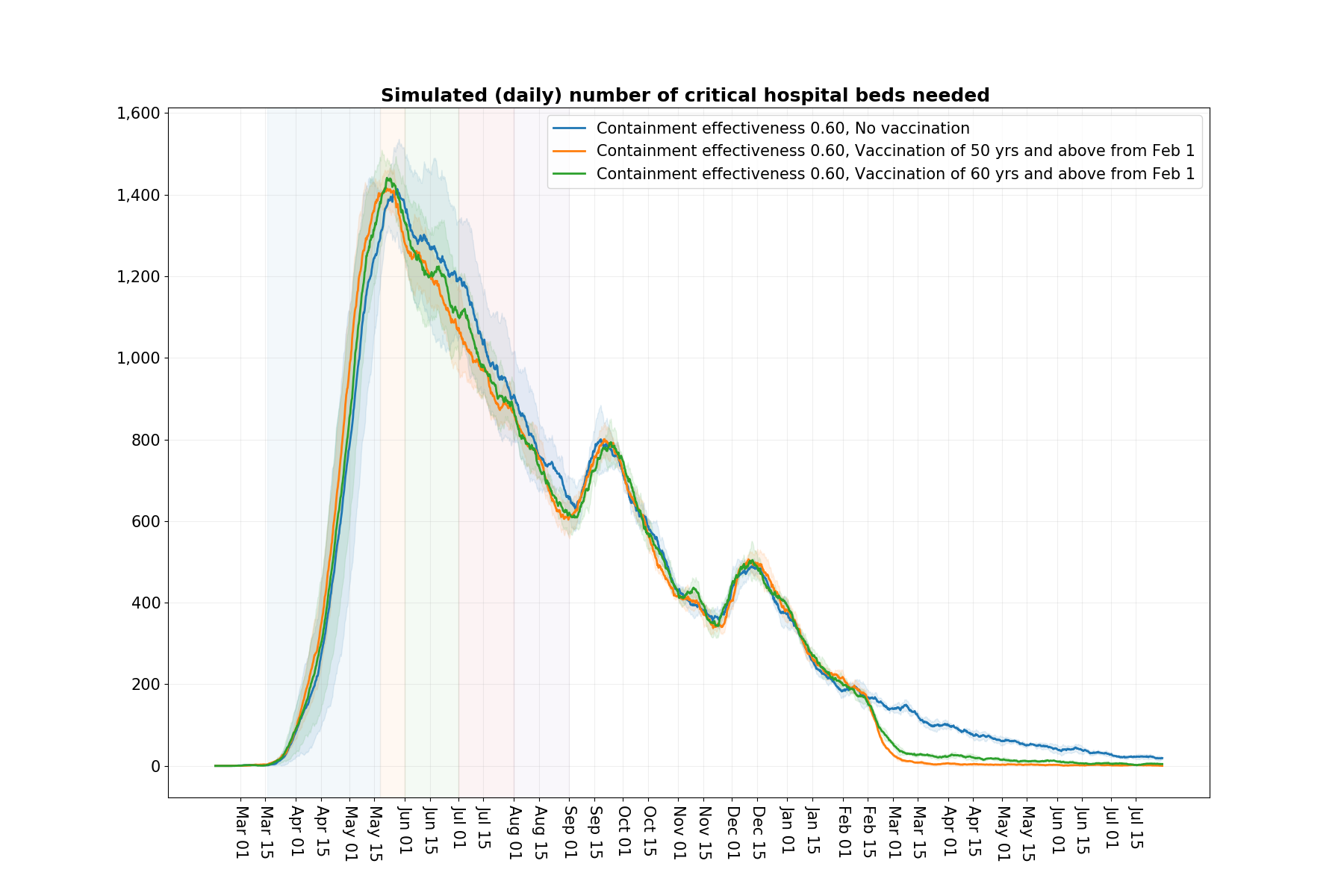}
      \caption{Simulated number of  daily hospitalised patients and critical patients
under scenarios of no vaccination, vaccinating people aged 50 years and above 
60 years and above on Feb 1.       
 Workplace opening schedule 
     is set at 5\% attendance, May 18
to May 31st, 15\% attendance in June, 25\% in July,
33\% in August, 50\% in September and October and fully open November onwards.
All the scenarios include the three festival relaxations. 
} \label{vaccination_figure_hospitalised}
 \end{figure}
 }

{
\begin{figure}
  \begin{center}
    \small
    \begin{tabular}{|c|c|c|}
      \hline
      {\bf Scenario} & {\bf Fatalities} & {\bf Hospitalisations (including critical cases)} \\
      \hline
      No Vaccination & 947 & 8840\\
      Vaccination of people aged 50 yrs and above & 340 & 2908\\
      Vaccination of people aged 60 yrs and above & 449 & 5342\\
      \hline
    \end{tabular}
	\end{center}
  \caption{Fatalities and hospitalisations (including critical cases) under scenarios of no vaccination, vaccination of people 50 years and above from Feb 1 and vaccination of people 60 years and above from Feb 1. Numbers shown are for a period of 6 months after Feb 1.    }
  \label{fatalities_vaccination}
\end{figure}
}

{
  \begin{figure}
      \centering
       \includegraphics[width=\linewidth]{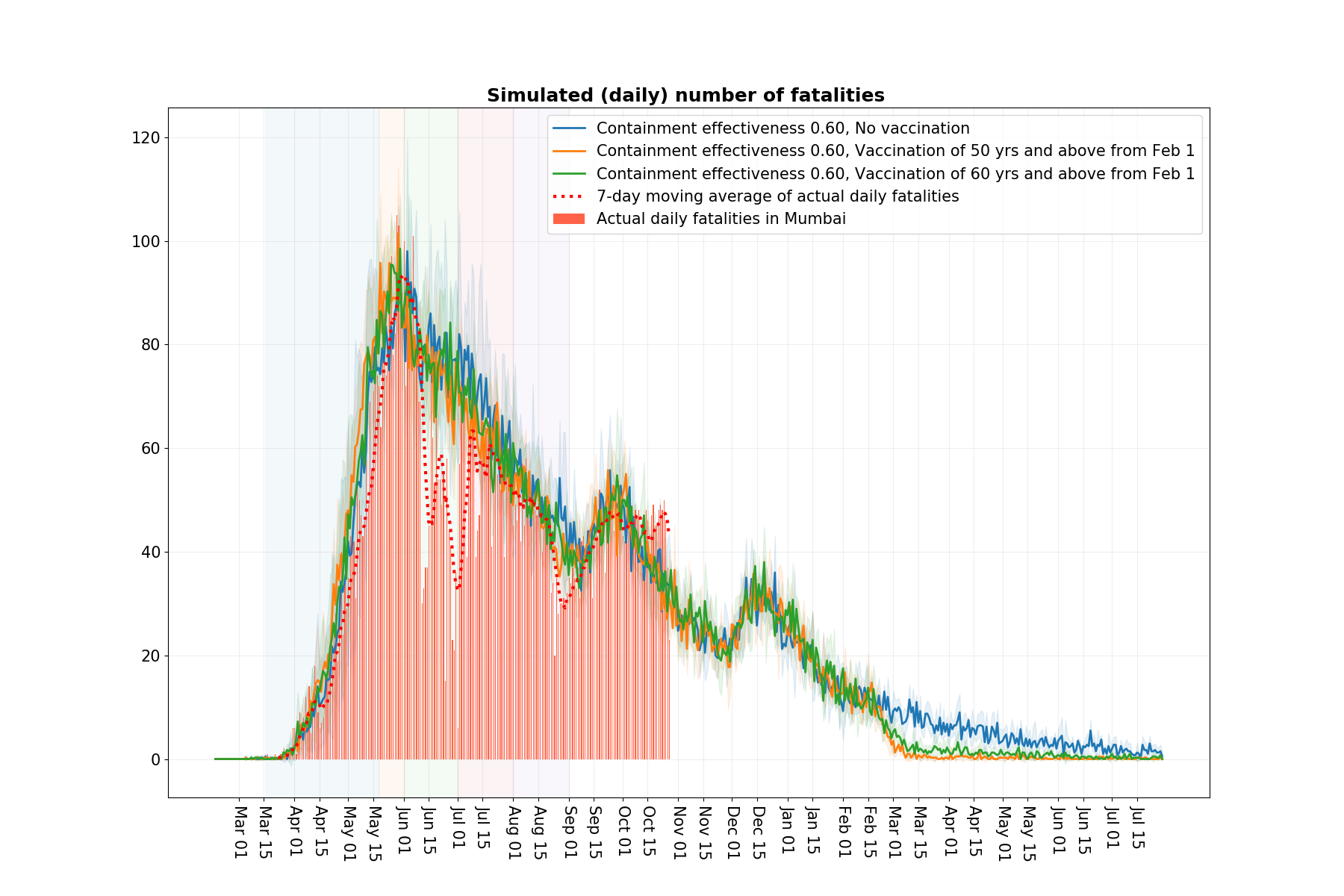}
      \caption{Simulated number of  daily fatalities under scenarios of no vaccination, vaccinating people aged 50 years and above, and  60 years and above on Feb 1.       
 Workplace opening schedule 
     is set at 5\% attendance, May 18
to May 31st, 15\% attendance in June, 25\% in July,
33\% in August, 50\% in September and October and fully open November onwards.  
All the scenarios include the three festival relaxations.
} \label{vaccination_figure_fatalities}
  \end{figure}
}

\section*{Acknowledgments}

We thank our colleague Piyush Srivastava for many useful suggestions that helped our analysis.  
We thank Piyush as well as our IISc collaborators
R. Sundaresan,  P. Patil, N. Rathod, A. Sarath, S. Sriram, and N. Vaidhiyan for their tireless efforts
in developing the IISc-TIFR Simulation model \cite{City_Simulator_IISc_TIFR_2020}
and their key role
in our earlier report on  Mumbai \cite{IISc_TIFR_2020_Mumbai_Report2}.

We thank Mrs. Ashwini Bhide, AMC, MCGM for her insights and for her crucial data inputs. We also thank Bhaskaran Raman for his critical inputs on the socio-economic effects of prolonged non-pharmaceutical interventions. We thank Siddarth Raman, a volunteer at BMC, for his help with data.

Authors acknowledge support of the Department of Atomic Energy, Government of India, to TIFR under project no. 12-R\&D-TFR-5.01-0500. RS is also supported by the Ramanujan Fellowship of DST.  
We also acknowledge the support of A.T.E. Chandra Foundation for this research.

\bibliographystyle{IEEEtran}
{
\bibliography{IEEEabrv,Mumbai_August}

\begin{thebibliography}{10}
\providecommand{\url}[1]{#1}
\csname url@samestyle\endcsname
\providecommand{\newblock}{\relax}
\providecommand{\bibinfo}[2]{#2}
\providecommand{\BIBentrySTDinterwordspacing}{\spaceskip=0pt\relax}
\providecommand{\BIBentryALTinterwordstretchfactor}{4}
\providecommand{\BIBentryALTinterwordspacing}{\spaceskip=\fontdimen2\font plus
\BIBentryALTinterwordstretchfactor\fontdimen3\font minus
  \fontdimen4\font\relax}
\providecommand{\BIBforeignlanguage}[2]{{%
\expandafter\ifx\csname l@#1\endcsname\relax
\typeout{** WARNING: IEEEtran.bst: No hyphenation pattern has been}%
\typeout{** loaded for the language `#1'. Using the pattern for}%
\typeout{** the default language instead.}%
\else
\language=\csname l@#1\endcsname
\fi
#2}}
\providecommand{\BIBdecl}{\relax}
\BIBdecl

\bibitem{maharashtra20:_easin}
{\relax Government of Maharashtra}, ``Easing of restrictions and phase-wise
  opening of lockdown. ({Mission Begin Again)},'' No.~DMU/2020/CR.~92/DisM-1,
  Aug. 2020, available at
  \url{https://twitter.com/CMOMaharashtra/status/1300420569863659522}.

\bibitem{Mumbai_sero2020}
\BIBentryALTinterwordspacing
A.~Malani, D.~Shah, G.~Kang, G.~N. Lobo, J.~Shastri, M.~Mohanan, R.~Jain, S.~T.
  Agrawal, S.~Juneja, S.~Imad, and U.~Kolthur-Seetharam, ``Seroprevalence of
  {SARS-CoV-2} in slums and non-slums of {M}umbai, india, during {J}une
  29--{J}uly 19, 2020,'' Aug. 2020. [Online]. Available:
  \url{https://www.medrxiv.org/content/10.1101/2020.08.27.20182741v1}
\BIBentrySTDinterwordspacing

\bibitem{City_Simulator_IISc_TIFR_2020}
\BIBentryALTinterwordspacing
S.~Agrawal, S.~Bhandari, A.~Bhattacharjee, A.~Deo, N.~Dixit, P.~Harsha,
  S.~Juneja, P.~Kesarwani, A.~Swamy, P.~Patil, N.~Rathod, R.~Saptharishi,
  S.~Shriram, P.~Srivastava, R.~Sundaresan, N.~K. Vaidhiyan, and S.~Yasodharan,
  ``\BIBforeignlanguage{en}{City-scale agent-based simulators for the study of
  non-pharmaceutical interventions in the context of the covid-19 epidemic},''
  Aug. 2020. [Online]. Available: \url{https://arxiv.org/abs/2008.04849}
\BIBentrySTDinterwordspacing

\bibitem{harsha2020covidmumbai}
P.~Harsha, S.~Juneja, and R.~Saptharishi, ``Covid-19 epidemic in mumbai: Long
  term projections, full economic opening, and containment zones versus contact
  tracing and testing,'' \emph{Preprint,
  \url{http://www.tcs.tifr.res.in/~sandeepj/avail_papers/Mumbai_September_Report.pdf}},
  2020.

\bibitem{malani2020serosurvey}
A.~Malani, D.~Shah, G.~Kang, G.~Lobo, J.~Shastri, M.~Mohanan, R.~Jain,
  S.~Agrawal, S.~Juneja, S.~Imad, and U.~Kolthur-Seetharam, ``{Seroprevalence
  of SARS-CoV-2 in slums and non-slums of Mumbai, India, during June 29-July
  19, 2020},'' \emph{medRxiv}, 2020.

\bibitem{mumbai_census2011_A}
``{District Census Handbook 2011 - Mumbai},''
  \url{https://censusindia.gov.in/2011census/dchb/DCHB_A/27/2723_PART_A_DCHB_MUMBAI.pdf},
  2014.

\bibitem{suburban_mumbai_census2011_A}
``{District Census Handbook 2011 - Suburban Mumbai},''
  \url{https://censusindia.gov.in/2011census/dchb/DCHB_A/27/2722_PART_A_DCHB_MUMBAI\%20SUBURBAN.pdf},
  2014.

\bibitem{who20:_q}
WHO, ``{Q\&A}: Considerations for the cleaning and disinfection of
  environmental surfaces in the context of covid-19 in non-health care
  settings,'' May 2020, available at
  \url{https://www.who.int/news-room/q-a-detail/q-a-considerations-for-the-cleaning-and-disinfection-of-environmental-surfaces-in-the-context-of-covid-19-in-non-health-care-settings}.

\bibitem{IISc_TIFR_2020_Mumbai_Report2}
\BIBentryALTinterwordspacing
P.~Harsha, S.~Juneja, P.~Patil, N.~Rathod, R.~Saptharishi, A.~Sarath,
  S.~Sriram, P.~Srivastava, R.~Sundaresan, and N.~Vaidhiyan,
  ``\BIBforeignlanguage{en}{Covid-19 epidemic study ii: Phased emergence from
  the lockdown in mumbai},'' Jun. 2020. [Online]. Available:
  \url{https://arxiv.org/abs/2006.03375}
\BIBentrySTDinterwordspacing

\bibitem{Laxminarayan2020}
\BIBentryALTinterwordspacing
R.~Laxminarayan, B.~Wahl, S.~R. Dudala, K.~Gopal, C.~Mohan, S.~Neelima, K.~S.
  Jawahar~Reddy, J.~Radhakrishnan, and J.~A. Lewnard, ``Epidemiology and
  transmission dynamics of {COVID-19} in two {I}ndian states,'' \emph{Science},
  2020. [Online]. Available:
  \url{https://science.sciencemag.org/content/early/2020/09/29/science.abd7672}
\BIBentrySTDinterwordspacing

\bibitem{WHO-pulse}
{World Health Organisation}, ``{Pulse survey on continuity of essential health
  services during the COVID-19 pandemic: interim report, 27 August 2020},''
  \url{https://www.who.int/publications/i/item/WHO-2019-nCoV-EHS_continuity-survey-2020.1},
  08 2020.

\bibitem{MCGM_census2011_report}
P.~H.~D. Municipal Corporation~of Greater~Mumbai, ``{Census 2011, FAQ
  Answers},''
  \url{https://portal.mcgm.gov.in/irj/go/km/docs/documents/MCGM\%20Department\%20List/Public\%20Health\%20Department/Docs/Census\%20FAQ\%20\%26\%20Answer.pdf}.

\bibitem{railway-survey}
M.~R. V.~C. Ltd., ``{Mumbai Sub-urban Rail Passenger Surveys and Analysis},''
  \url{https://mrvc.indianrailways.gov.in/works/uploads/File/ExecutiveSummarywilber\%20FINAL.pdf},
  2013.

\bibitem{verity2020estimates}
R.~Verity, L.~C. Okell, I.~Dorigatti, P.~Winskill, C.~Whittaker, N.~Imai,
  G.~Cuomo-Dannenburg, H.~Thompson, P.~G. Walker, H.~Fu \emph{et~al.},
  ``Estimates of the severity of coronavirus disease 2019: a model-based
  analysis,'' \emph{The Lancet Infectious Diseases}, 2020.

\bibitem{BMCdashboard}
``{BMC COVID-19 Response War Room Dashboard},''
  \url{http://stopcoronavirus.mcgm.gov.in/assets/docs/Dashboard.pdf}, 04 2020.

\bibitem{ferguson2020report}
N.~Ferguson, D.~Laydon, G.~Nedjati~Gilani, N.~Imai, K.~Ainslie, M.~Baguelin,
  S.~Bhatia, A.~Boonyasiri, Z.~Cucunuba~Perez, G.~Cuomo-Dannenburg
  \emph{et~al.}, ``{Report 9: Impact of non-pharmaceutical interventions (NPIs)
  to reduce COVID19 mortality and healthcare demand},'' \emph{Tech. Report},
  2020.

\bibitem{TOI_article}
``{Maharashtra stop testing dead bodies to ease painful delays for kin},''
  \url{https://timesofindia.indiatimes.com/city/mumbai/maharashtra-stops-testing-bodies-says-will-ease-painful-delays-for-kin/articleshow/76474747.cms},
  06 2020.

\end{thebibliography}
}

\end{document}